\documentclass{aastex63}
\usepackage{comment}
\usepackage{lineno}
\shorttitle{Velocity dispersion profile and rotation curve of NGC 1904}
\shortauthors{Leanza et al.}

\graphicspath{{./}{figures/}}

\begin{document}

\title{The ESO-VLT MIKiS survey reloaded: velocity dispersion profile
  and rotation curve of NGC 1904\footnote{Based on observations
  collected at the European Southern Observatory, Cerro Paranal
  (Chile), under proposal 193.D-0232 (PI: Ferraro) and 0104.D-0636
  (PI: Ferraro) }}

\correspondingauthor{Silvia Leanza}
\email{silvia.leanza2@unibo.it}

\author[0000-0001-9545-5291]{Silvia Leanza}
\affil{Dipartimento di Fisica e Astronomia, Universit\`a di Bologna, Via Gobetti 93/2 I-40129 Bologna, Italy}
\affil{INAF-Osservatorio di Astrofisica e Scienze dello Spazio di Bologna, Via Gobetti 93/3 I-40129 Bologna, Italy}

\author[0000-0002-7104-2107]{Cristina Pallanca}
\affil{Dipartimento di Fisica e Astronomia, Universit\`a di Bologna, Via Gobetti 93/2 I-40129 Bologna, Italy}
\affil{INAF-Osservatorio di Astrofisica e Scienze dello Spazio di Bologna, Via Gobetti 93/3 I-40129 Bologna, Italy}

\author[0000-0002-2165-8528]{Francesco R. Ferraro}
\affil{Dipartimento di Fisica e Astronomia, Universit\`a di Bologna, Via Gobetti 93/2 I-40129 Bologna, Italy}
\affil{INAF-Osservatorio di Astrofisica e Scienze dello Spazio di Bologna, Via Gobetti 93/3 I-40129 Bologna, Italy}

\author[0000-0001-5613-4938]{Barbara Lanzoni}
\affil{Dipartimento di Fisica e Astronomia, Universit\`a di Bologna, Via Gobetti 93/2 I-40129 Bologna, Italy}
\affil{INAF-Osservatorio di Astrofisica e Scienze dello Spazio di Bologna, Via Gobetti 93/3 I-40129 Bologna, Italy}

\author[0000-0003-4237-4601]{Emanuele Dalessandro}
\affil{INAF-Osservatorio di Astrofisica e Scienze dello Spazio di Bologna, Via Gobetti 93/3 I-40129 Bologna, Italy}

\author{Livia Origlia}
\affil{INAF-Osservatorio di Astrofisica e Scienze dello Spazio di Bologna, Via Gobetti 93/3 I-40129 Bologna, Italy}

\author{Alessio Mucciarelli}
\affil{Dipartimento di Fisica e Astronomia, Universit\`a di Bologna, Via Gobetti 93/2 I-40129 Bologna, Italy}
\affil{INAF-Osservatorio di Astrofisica e Scienze dello Spazio di Bologna, Via Gobetti 93/3 I-40129 Bologna, Italy}

\author[0000-0002-6092-7145]{Elena Valenti}
\affil{European Southern Observatory, Karl-Schwarzschild-Strasse 2, 85748 Garching bei Munchen, Germany}
\affil{Excellence Cluster ORIGINS, Boltzmann\--Stra\ss e 2, D\--85748 Garching bei M\"{u}nchen, Germany }

\author{Maria Tiongco}
\affil{University of Colorado, JILA and Department of Astrophysical and Planetary Sciences, 440 UCB, Boulder, CO 80309 - USA}

\author{Anna Lisa Varri}
\affil{Institute for Astronomy, University of Edinburgh, Royal Observatory, Blackford Hill, Edinburgh EH9 3HJ, UK}
\affil{School of Mathematics, University of Edinburgh, Kings Buildings, Edinburgh EH9 3FD, UK}

\author[0000-0003-2742-6872]{Enrico Vesperini}
\affil{Department of Astronomy, Indiana University, Bloomington, IN, 47401, USA}




\begin{abstract}

We present an investigation of the internal kinematic 
properties of M79 (NGC 1904). Our study is based on radial velocity 
measurements obtained from the ESO-VLT Multi-Instrument 
Kinematic Survey (MIKiS) of Galactic globular clusters for more than 1700 individual stars distributed between $\sim 0.3\arcsec$ and $770\arcsec$ ($\sim14$ three-dimensional half-mass radii), from the center.
Our analysis reveals the presence of ordered line-of-sight rotation with a rotation axis almost aligned along the East-West direction and a velocity peak of $1.5$ km s$^{-1}$ at $\sim 70\arcsec$ from the rotation axis.
The velocity dispersion profile is well described by the same King
model that best fits the projected density distribution, 
with a constant central plateau at $\sigma_0\sim 6$ km s$^{-1}$.
To investigate the cluster rotation in the plane of the sky, 
we have analyzed the proper motions provided by the Gaia EDR3, 
finding a signature of rotation with a maximum amplitude of $\sim 2.0$ km s$^{-1}$ at $\sim 80\arcsec$ from the cluster center.
Analyzing the three-dimensional velocity distribution, for a sub-sample of 130 stars, we confirm the presence of systemic rotation and find a rotation axis inclination angle of 37\arcdeg \ with respect to the line-of-sight.
As a final result, the comparison of the observed rotation curves with the results of 
a representative {\it N}-body simulation of a rotating star cluster 
shows that the present-day kinematic properties of NGC 1904 
are consistent with those of a dynamically old system that has 
lost a significant fraction of its initial angular momentum.

\end{abstract}



\section{Introduction} \label{sec:intro}
Galactic Globular clusters (GGCs) are the most populous and oldest
stellar systems where stars can be individually observed. Moreover,
they are collisional systems, where the frequent gravitational
interactions among stars can make the characteristic timescale for
dynamical evolution significantly shorter than their age, depending on
intrinsic properties at formation (e.g., total mass, central mass
density, binary content, etc.) and external effects in the environment
they are embedded in (e.g., Galactic tides, disk shocks). In fact, in
spite of their similar chronological ages ($\sim 12$ Gyr;
\citealp{forbes+10}), GGCs show different stages of internal dynamical
evolution (see \citealp{ferraro+20} and reference therein) and
therefore are ideal laboratories where the complex interplay between
stellar population properties and dynamical evolutionary effects can
be empirically investigated.  The innermost core regions of GGCs are
expected to offer the ideal environment for the occurrence of stellar
interactions able to generate exotic objects, like interacting
binaries, blue stragglers, millisecond pulsars \citep{bailyn95,
  pooley+03, ransom+05, ferraro+97, ferraro+03, ferraro+18a}, and even
the long sought class of intermediate-mass black holes (IMBHs;
e.g. \citealp{giersz+15}). Indeed, the extrapolation of the
``Magorrian relation'' \citep{magorrian+98} down to the IMBH mass
scale naturally leads to the mass regime of GGCs for the hosting
stellar system. In addition, numerical simulations
\citep[e.g.][]{portegies+04, freitag+07, giersz+15} confirm that GGCs
are ideal habitats for the formation of IMBHs.  The accurate
characterization of GGCs in terms of their structural properties,
internal kinematics and dynamical status is a crucial step for the
proper understanding of how dynamical processes affect the
evolutionary history of these systems and impact the formation of
stellar exotica.

Our group is addressing this problem by combining a variety of 
complementary perspectives: 
{\it (i)} by constructing a new generation of high quality star density
profiles derived from star counts instead of surface brightness (see
\citealp{miocchi+13, lanzoni+07a, lanzoni+10, lanzoni+19,
  pallanca+21}); {\it (ii)} by investigating the population of stellar
exotica (\citealt{ferraro+01, ferraro+03, ferraro+15, ferraro+16,
  pallanca+10, pallanca+13, pallanca+14, pallanca+17, cadelano+17,
  cadelano+18, cadelano+20}) and their connection with the dynamical
evolution of the parent cluster (see \citealp{ferraro+09, ferraro+12,
  ferraro+18a, ferraro+19, lanzoni+16}); {\it (iii)} by characterising
the three-dimensional global velocity space through the analysis
of the velocity dispersion profile and rotation
curve from resolved star
spectroscopy \citep{lanzoni+13, lanzoni+18a, lanzoni+18b, ferraro+18b}
and proper motions (see \citealt{raso+20}). The determination of GGC
internal kinematics from resolved star velocities is particular
relevant and challenging. In this context we promoted the ESO-VLT
Multi-Instrument Kinematic Survey (hereafter the MIKiS survey;
\citealt{ferraro+18b, ferraro+18c}), a project specifically designed
to characterize the kinematical properties of a sample of GGCs in
different dynamical evolutionary stages from the radial
velocities (RVs) of hundred individual stars distributed over the
entire radial range of each stellar system. To this end, the survey
fully exploits the spectroscopic capabilities of different instruments
currently available at the ESO Very Large Telescope (VLT): originally
designed to use the adaptive-optics (AO) assisted integral-field
spectrograph SINFONI, the multi-object integral-field spectrograph
KMOS, and the multi-object fiber-fed spectrograph FLAMES/GIRAFFE, it
has been recently complemented with individual projects and an ongoing
large program (PI: Ferraro) fully exploiting the remarkable
performances of the AO-assisted integral-field spectrograph MUSE.

In this paper, we present the velocity dispersion profile and rotation
curve of NGC 1904, a well-known metal-intermediate and low-extinction
GGC, with [Fe/H]$=-1.6$ dex, E$(B-V)=0.01$ \citep{harris96, ferraro+99},
and a blue extended Horizontal Branch \citep{ferraro+92, lanzoni+07,
  dalessandro+13}.  This cluster is particularly intriguing since it
was indicated as the possible host of an IMBH of $\sim 3000
M_\odot$ on the basis of the shape of the velocity dispersion profile
obtained from integrated-light spectra in the innermost regions of the
system \citep{lutz+13}. In \citet{ferraro+18b} we presented the line
of sight kinematics of the external region of the system as obtained
from individual RV measurements. Here we complement those data with
recent MUSE observations of the innermost $\sim 15\arcsec$ in the
highest available spatial resolution configuration, and with archive
MUSE data at lower spatial resolution extending to radial distances of
$ \sim 67\arcsec$ from the center.  

The paper is organized as follows. Section \ref{sec:obs} provides the description of the observations and
the adopted data reduction procedures. 
In Section \ref{sec:analysis} we discuss the selection of the samples,
the determination of the stellar RVs and the
homogenization of the different datasets. 
The details on the kinematic analysis of the line-of-sight velocities and the derived results are presented in Section \ref{sec:result}, together with the results obtained from the analysis of the proper motions from Gaia EDR3. The conclusions are then presented in Section \ref{sec_discuss}.

\section{Observations and data reduction}
\label{sec:obs}
As discussed above, to construct the velocity dispersion profile of
NGC 1904 we complemented the catalog presented in \citet{ferraro+18b}
with a new set of spectroscopic data at high spatial resolution. To
acquire spectra of individual stars in the innermost regions of NGC
1904, we took advantage of the superb spatial resolution capabilities
of the AO-assisted integral-field spectrograph MUSE in the Narrow
Field Mode (NFM) configuration \citep{10.1117/12.856027}. MUSE is
mounted on the Yepun, the VLT-UT4 telescope at the ESO Paranal
Observatory, and it is equipped with
the Adaptive Optics Facility (AOF) of the VLT and the GALACSI-AO
module. It has a modular structure composed of 24 identical Integral
Field Units (IFUs) and, in the nominal mode\footnote{MUSE provide two instrument
mode for the wavelength coverage: nominal and extended mode, which correspond 
to $4800 - 9300$ \AA\ and $4650 - 9300$ \AA, respectively.},
it samples the wavelength range $4800 - 9300$
\AA\ with a spectral resolution R $\sim3000$ at $\lambda\sim8700$
\AA.

The NFM dataset acquired in NGC 1904 consists of a mosaic of seven
MUSE/NFM pointings covering the innermost $\sim15\arcsec$ from the
cluster centre \citep{lanzoni+07}, each pointing having a field of
view of $7.5\arcsec\times 7.5\arcsec$ and spatial sampling of
$0.025\arcsec$/pixel. The observations have been collected on 2019,
November 1-5 and December 2-6 (ESO proposal ID: 0104.D-0636(A), PI:
Ferraro, see Table \ref{tab:data}). In general, for each pointing 3 exposures were
acquired following a small dithering pattern (smaller than
$1.0\arcsec$).  A $90^\circ$ rotation of the detector has been set
between consecutive exposures, in order to correct for possible
systematic effects of individual spectrographs, to improve the flat
fielding quality and, hence, to reach a homogeneous image quality
across the entire field of view. A series of 835 s long exposures were
secured under good seeing conditions: the average DIMM seeing during
the observations was always better than $\sim0.7\arcsec$.

The NFM dataset has been complemented with archive observations (ESO
program ID: 098.D-0148(A), PI: Dreizler) acquired with MUSE in the
Wide Field Mode (WFM) configuration, providing velocity measures at
intermediate distances from the cluster center and allowing a proper
overlap with the data presented in \citet{ferraro+18b}.  A mosaic of 4
WFM pointings has been included in the analysis, each pointing having
an exposure time of 120 s and providing a $1\arcmin\times 1\arcmin$
field of view, with a spatial sampling of $0.2\arcsec$/pixel. This
dataset covers a cluster region up to $\sim 67\arcsec$ from the
center, including the area sampled by the high-resolution NFM
observations.  The wavelength range, the spectral resolution and the
adopted strategy for the acquisition of the images are the same
previously described for NFM observations.

\begin{figure}[ht!]
\centering
\includegraphics[width=18cm, height=8cm]{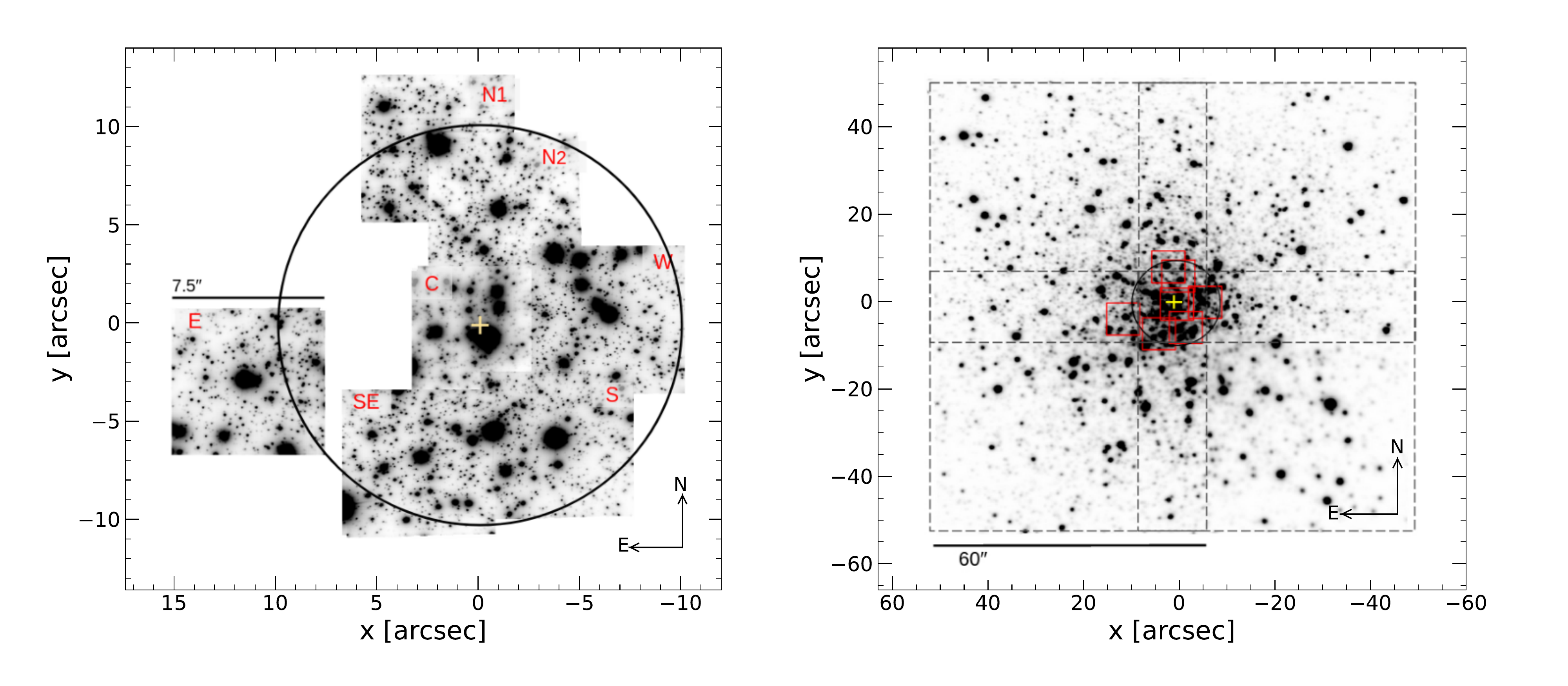}
\caption{{\it Left:} Reconstructed $I$-band images of the seven MUSE/NFM
  pointings. The x and y axes represent the projected positions of the stars with respect to the cluster center (yellow cross, from \citealp{lanzoni+07}). The circle is centered on the center and has a radius of $10\arcsec$.
  {\it Right:} Reconstructed $I$-band image of the four MUSE/WFM
  pointings (each sampling $60\arcsec\times 60\arcsec$ on the sky;
  dashed squares), with the location of the seven NFM pointings
  overplotted in red. The yellow cross and the circle are as in the
  left panel.}
  \label{fig:puntamenti}
\end{figure}

The MUSE data reduction (for both NFM and WFM images) was performed
with the dedicated ESO pipeline \citep{weilbacher2020data}. It
consists of two main steps. The first one performs the basic reduction
of individual IFUs, including bias subtraction, flat fielding, and
wavelength calibration. The second step transforms pre-processed data into physical quantities, by performing the flux calibration, sky subtraction and astrometric calibration for each IFU and applying the heliocentric velocity correction to all data.
Then the data from all 24 IFUs are combined into a single three-dimensional
data cube. Finally, the pipeline combines the data cubes of each
individual exposure into a single final data cube for every pointing,
taking into account possible offsets and rotations among different
exposures. Figure \ref{fig:puntamenti} (left panel) shows the
reconstructed data cube image of the NFM observations for the 7
available pointings named according to their position with respect to
the cluster center: \textit{Center} (C), \textit{South} (S),
\textit{East} (E), \textit{West} (W), \textit{North} (N1), \textit{North} (N2)
\textit{South-East} (SE). Note that only two exposures 
were acquired in fields N2 and E because of technical problems. The
reconstructed image of the four WFM pointings is shown in the right
panel of the same figure.

Indeed, already from a first visual inspection, the NFM observations
immediately appears of superb quality, well comparable to the quality
of HST images.  This can be even better appreciate in Figure
\ref{fig:test}, where the reconstructed NFM image of the W pointing
(central panel) is compared with an image of same region as seen from
the Planetary Camera of the HST/WFPC2
(right panel), and as sampled by the WFM observations (left panel).
From this comparison, it is clear that NFM observations are mandatory
to obtain a large number of RV measurements from individual stars in
the core of the cluster.

\begin{deluxetable*}{cCClcC}
\tablecaption{MUSE/NFM dataset.}
\tablewidth{0pt}
\tablehead{
\colhead{ Name } & \colhead{ RA }  & \colhead{Dec}  &  \colhead{}  &
\colhead{Date} & \colhead{N$_{\rm exp}$}  
}
\startdata
C & 81.045674 & -24.52522 & &2019-11-05 & 3 \\
S & 81.044944 & -24.52612 & & 2019-12-03 & 3 \\
SE & 81.046915 & -24.52657 & & 2019-12-04 & 3 \\
 &             &            & & 2019-12-06 & 3 \\
E  & 81.049125 & -24.52575 & & 2019-12-03 & 2 \\
W & 81.043994 & -24.52513 & & 2019-12-04 & 3 \\
N1 &  81.046829 & -24.52267 & & 2019-12-06 & 3\\
N2 &  81.046788 & -24.52191 & & 2019-11-02 & 2\\
\enddata
\tablecomments{Name, coordinates (in degree), observation date and
  number of exposures (N$_{\rm exp}$) for each NFM pointing
  analyzed in this paper.}
\label{tab:data}
\end{deluxetable*}

\section{Analysis}
\label{sec:analysis}
\subsection{Photometric analysis and selection of target stars}
\label{sec:selection}
The photometric analysis has been performed on the two-dimensional
image extracted from each data cube from the stacking of MUSE slices in the wavelength
range $8540-8550$\AA\ (which is the region of the Calcium triplet,
providing the highest signal-to-noise ratio, S/N). To determine the position of the centroid of individual
sources and identify possible blends in the MUSE data, we took
advantage of catalogs obtained from HST observations, which guarantee
the necessary angular resolution to properly resolve stars in the
highly crowded central regions of the system.
For the innermost area we used a catalog obtained from images acquired
with the HST/WFPC2 Planetary Camera (having a pixel-scale of
$0.046\arcsec$ pixel$^{-1}$), in the filters F439W (hereafter $B$) and
F555W (hereafter $V$).  The catalog has been placed on the absolute
astrometric system through cross-correlation with the photometric
catalog of \citet{10.1093/mnras/stz585}\footnote{For the photometric
catalog, see
\url{https://www.canfar.net/storage/list/STETSON/homogeneous/MNRAS_Photometry_for_48_Clusters/}},
and the instrumental magnitudes were then calibrated using the catalog
discussed in \citet{lanzoni+07}. 

\begin{figure}[ht!]
\centering
\includegraphics[width=16.4cm, height=5cm]{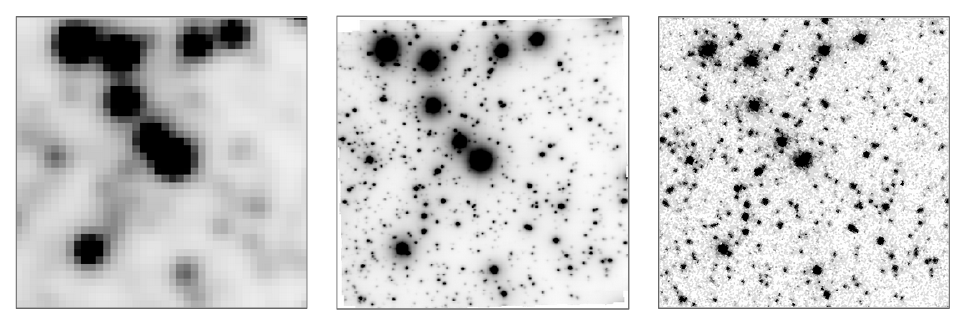}
\centering
\caption{Reconstructed MUSE/NFM image of the W pointing (central
  panel), compared with that obtained with the Planetary Camera of the
  HST/WFPC2 (right panel) and the MUSE/WFM observations (left
  panel). In all cases the field of view samples an area of $\sim
  8\arcsec\times 8\arcsec$ on the sky.}
\label{fig:test}
\end{figure}  

A quick photometric analysis of the NFM images provided us with the
preliminary position of the brightest sources.  Then, the
cross-correlation with the HST catalog allowed us to assign to every
star observed with HST an (X,Y) position in the coordinate system of
the NFM images. The best-fit PSF model has been determined by
analyzing the brightest $\sim 10$ stars in each pointing with
daophot/psf \citep{Stetson_1987}. Finally, we ran daophot/allstar
\citep{Stetson_1987} to perform the PSF fitting photometry at the
(X,Y) centroid position of all the HST sources with
$V<21$.\footnote{This magnitude cut has been set after the preliminary
photometric analysis revealed that sources fainter than this limit
are not reliably detected by the available NFM observations.}  As a
result, the PSF fitting analysis yields, for each image, the accurate
position of the centroid and the instrumental magnitudes of each
stellar source.  As a double check, we verified that the position of the
stellar centroids obtained from the PSF fitting procedure was not
significantly altered by the procedure (this could happen, in
principle, for the effect of a bright neighbor, or in the case of
blending). Stars with centroid deviations larger than $\delta X$ and $\delta Y>0.5\arcsec$, with respect the HST position, have been excluded from the analysis. As final result of the entire
procedure, for each star identified in each MUSE pointing, we obtain
the $X$ and $Y$ positions in the datacube, the RA, Dec absolute
coordinates, and the $B$ and $V$ magnitudes.

A similar analysis has been performed for the WFM dataset. In this
case, the HST reference catalog used to identify the stellar centroids
has been obtained from the combination of the two WFPC2 datasets
presented in \citet{lanzoni+07} and different magnitude cuts have
been adopted: $V<17$ for $r<20\arcsec$, and $V <19$ in the region
between $20\arcsec$ and $80\arcsec$ from the center.  The adopted limits are
brighter than those used in the analysis of the NFM data because of
the brighter magnitude level reached by the WFM observations, which is
clearly apparent from the comparison shown in Figure \ref{fig:test}
and is due to shorter exposure times.  Moreover, these limits depend
on the radial distance from the center because, due to the low spatial
resolution of the WFM observations, only brighter stars can be
resolved in the higher density regions.

\begin{figure}[ht!]
\centering
\includegraphics[width=17cm, height=10cm]{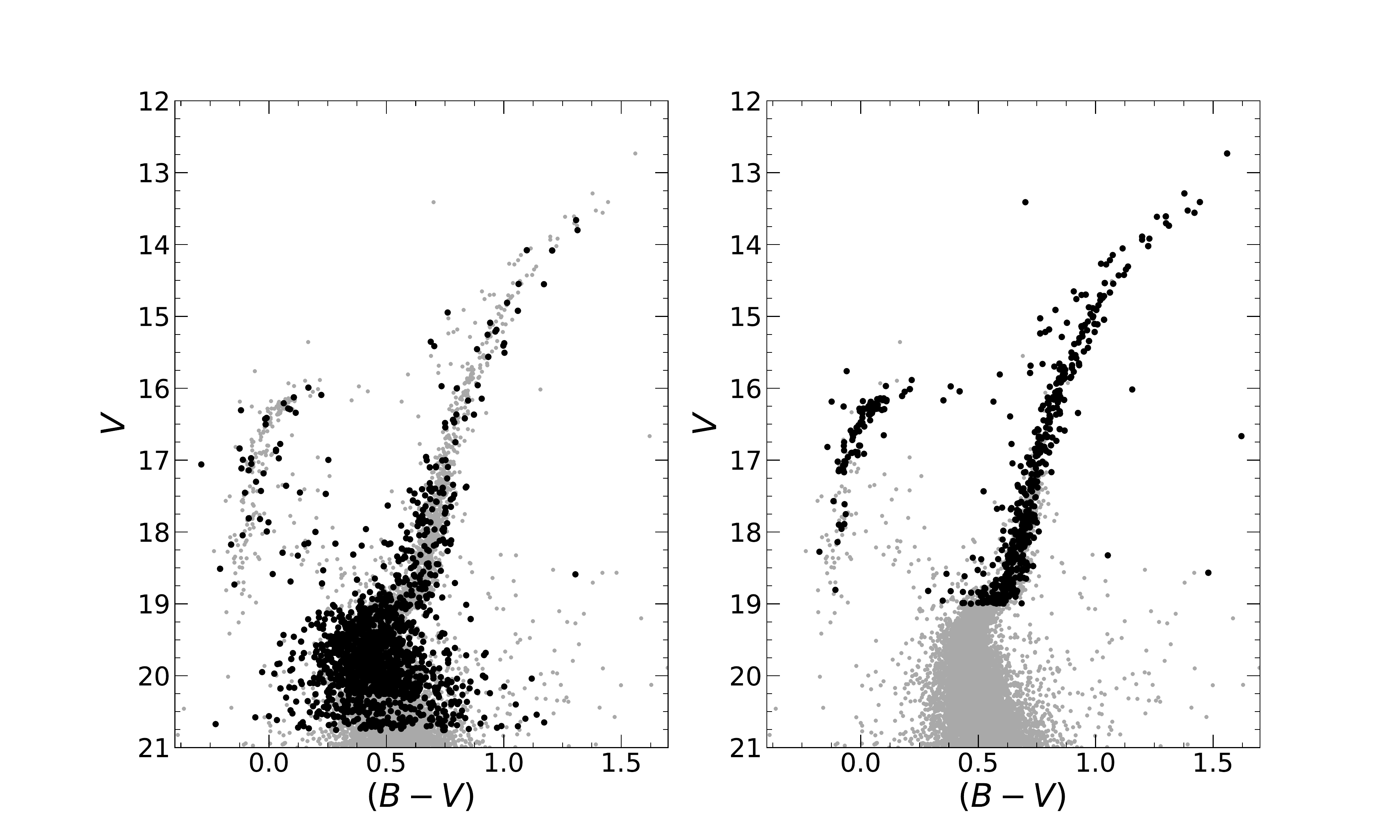}
\caption{CMD of NGC 1904 (gray dots) obtained from the WFPC2 catalogs
  presented in \cite{lanzoni+07}, with all the stars identified in
  the photometric analysis of the MUSE NFM and WFM images highlighted
  as large black dots (left and right panels, respectively).}
\label{fig:cmd_pre}
\end{figure}

The ($V, B-V$) color-magnitude diagrams (CMDs) of all the stars
identified in the MUSE NFM and WFM images are overplotted to the CMD
of NGC 1904 \citep[from][]{lanzoni+07} in the left and right panels
of Figure \ref{fig:cmd_pre}, respectively.

\subsection{Stellar Radial Velocities}
\label{sec:rv}
\begin{figure}[ht!]
\centering \includegraphics[width=14cm, height=7.2cm]{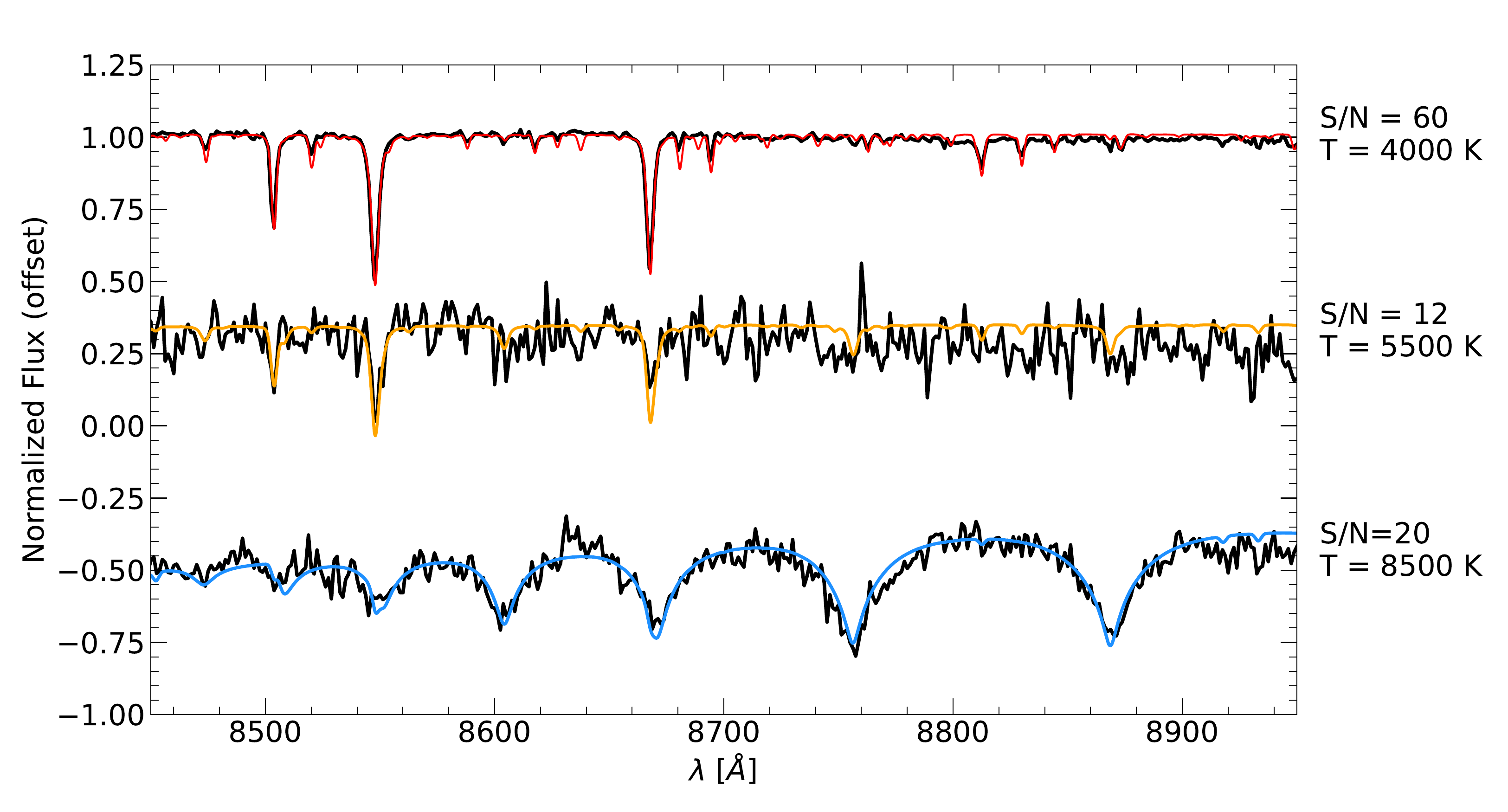}
\caption{Examples of spectra acquired with MUSE/NFM in the region of
  the Calcium triplet lines (in black) for different S/N values and
  for stars with different atmospheric parameters (see labels). The
  best-fit template spectra, shifted by the adopted RV value, are
  overplotted as colored lines.  In the bottom spectrum, corresponding
  to an HB star with effective temperature $T=8500$ K, the Calcium
  triplet lines is strongly blended with the hydrogen Paschen lines. The examples shown are a good representation also of the spectra acquired with MUSE/WFM.}
\label{fig:spettri}
\end{figure}

As mentioned in the Introduction, our approach to determine the
internal kinematics of GGCs consists in measuring the RVs of resolved
sources from individual stellar spectra.  The procedure adopted for
each MUSE data cube can be summarized as follows:
\begin{itemize}
\item{\it step (1): extraction of the spectra.} The spectra have been
  extracted at the coordinates of the stellar centroids identified during
  the photometric procedure (see Section \ref{sec:selection}). Since
  the targets are located in highly crowded regions, in order to limit the possible contamination from close sources, we made the conservative choice of extracting only the spectrum of the centroid spaxel of the target star and the 4 adjacent ones along the X and Y directions 
  (hence, the 5 spaxels drawing a
  cross centered on the centroid of the source).
\item{\it step (2): normalization of the spectra.} The spectra have
  been normalized to the continuum estimated by a spline fitting of
  the portion of the spectrum in the wavelength range $6700 - 9300$ \AA. The S/N has been estimated for each
  spectrum as the ratio between the average of the counts and its
  standard deviation in the wavelength range $8000 - 9000$ \AA.
\item{\it step (3): selection of template spectra.} A library of
  template synthetic spectra has been computed with the SYNTHE code
  (\citealt{sbordone2004atlas} and \citealt{2005MSAIS...8...14K}),
  adopting the cluster metallicity (\citealt{harris96}) and
  appropriate atmospheric parameters (effective temperature and
  gravity) according to the target evolutionary stage derived from the
  CMD.
\item {\it step (4): measurement of RVs.} The RV of almost all the
  selected stars has been measured from the Doppler shift of the Calcium
  Triplet lines in the normalized spectra, in the wavelength range
  8450 - 8740 \AA. In the case of horizontal branch (HB) stars,
  because of the blending between the Calcium triplet lines and the
  hydrogen Paschen lines, we used the hydrogen lines in the spectral
  range 8500 - 8930 \AA.  The adopted method consists in the analysis
  of the residuals between the observed spectrum and a set of
  reference spectra shifted in wavelength by quantities corresponding
  to different velocities, testing all the values in the range $0 - 400$
  km s$^{-1}$, at 0.1 km s$^{-1}$ steps. The adopted RV is obtained
  from the wavelength shift that minimizes the standard deviation of
  the residuals.
\end{itemize}
Figure \ref{fig:spettri} shows an example of the output of this
procedure. The observed spectra, with different S/N and corresponding
to stars with different temperatures, are shown in black. For each of
them, the figure also shows the best-fit template shifted by the
adopted value of RV.  As discussed above, while the Calcium triplet
has been used to measure RV of the two colder stars ($T<5000$ K, two
top spectra), the hydrogen Paschen lines have been exploited for the
hottest source ($T=8500$ K, bottom spectrum).  The RV values obtained
from the adopted method have been compared with the results of the
``standard'' cross-correlation approach
(\citealt{1979AJ.....84.1511T}) implemented in IRAF, always showing
good agreement independently of the spectrum S/N.

\begin{figure}[ht!]
\centering
\includegraphics[width=17cm, height=8.6 cm]{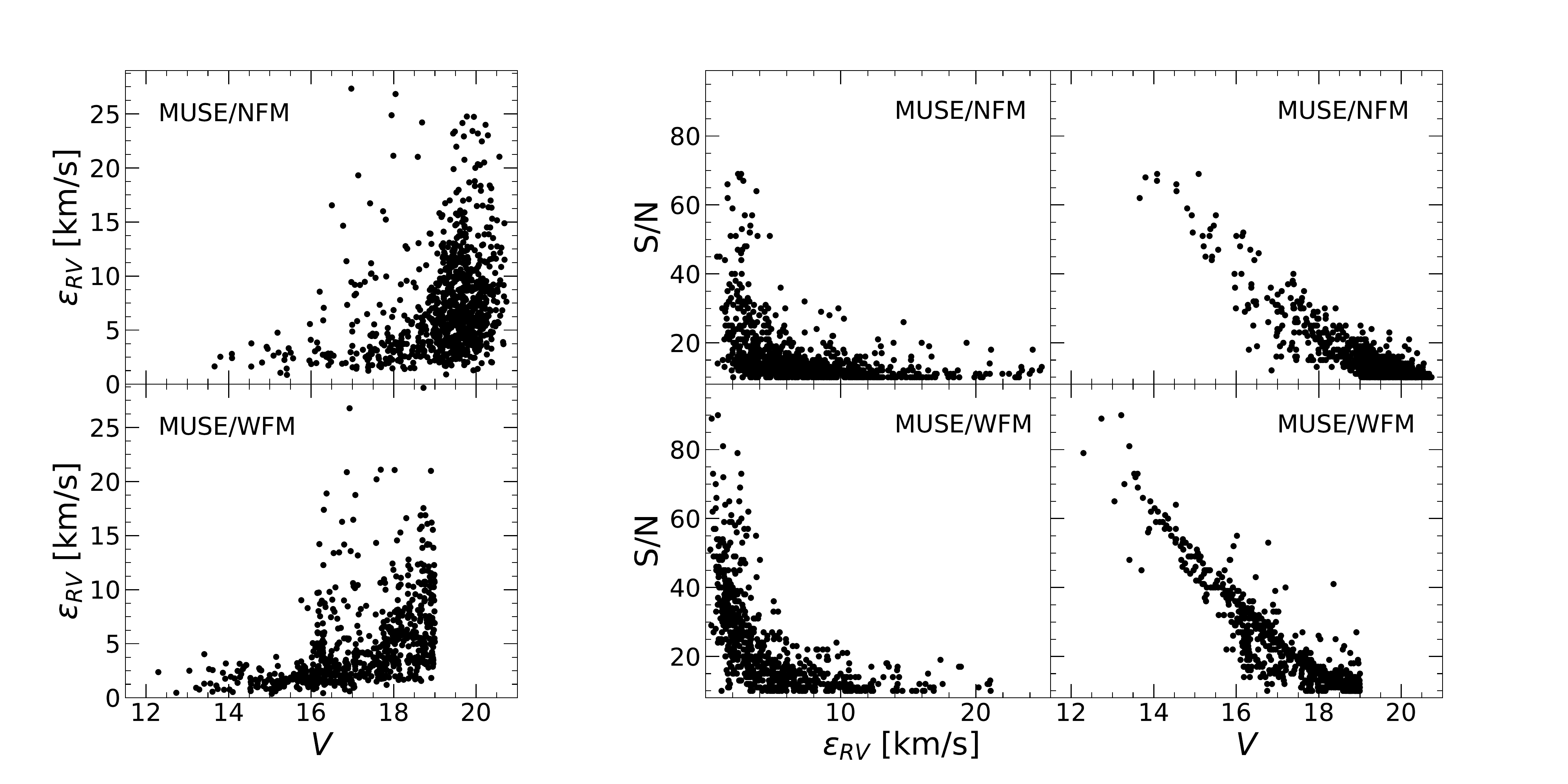}
\caption{{\it Left panel:} RV uncertainty ($\epsilon_{RV}$) as a function of the star magnitude for the targets of the MUSE/NFM and MUSE/WFM samples (top and bottom panel, respectively). {\it Central and right panels}: S/N as a function of velocity error and star magnitude, for the NFM targets (top panels) and the WFM sample (bottom panels).}
\label{fig:snr}
\end{figure}

The final value of RV for each star and its uncertainty have been then computed, respectively, as the weighted mean and the weighted standard deviation of the measures obtained from the 5 extracted spaxels, after a $3\sigma$-rejection procedure aimed at removing clearly discrepant values generated by spurious effects.
The relative weights of the five
measures have been defined according to the fraction of the star light
sampled by each spaxel, estimated from the adopted PSF model: we
assumed weight $=1$ for the central and most exposed spaxel, while
weight $=0.4$ and 0.5 for the 4 adjacent spaxels in the NFM and WFM
samples, respectively. 
We used the sub-samples of stars having multiple measures to check for the reliability of the RV uncertainties.  By applying the method described in Section 4.1 of \citet{kamann+16}, we found that the adopted errors could be slightly overestimated (by $\sim 0.4$ km s$^{-1}$). However, by taking into account that the sample of repeated measures is poor ($\sim 80 - 100$ stars) and the impact on the results of the work is negligible, we decided to introduce no corrective factors.
The final S/N associated to each star is the
weighted average of the S/N values of the considered spectra. 
For the following analysis, we finally selected
only those stars for which the RV value was determined from at least 3
spaxels over 5, and with S/N$\geq 10$. The typical uncertainties are
$<5.0$ km s$^{-1}$ for the brightest stars ($V<16$), while they
increase for fainter magnitudes according to the corresponding decrease in S/N (see Figure \ref{fig:snr}). To produce a homogeneous
final catalog, we first checked for possible systematic offsets in RV
among the different MUSE pointings. To this purpose, we compared the
RV values of the stars in common between two overlapping fields,
always finding good agreement within the errors (only the E pointing
has no objects in common with the other fields). In the cases of
multiple exposures for the same star, we determined its final RV as
the weighted mean of all the measures, by using the individual errors
as weights.


To sample the entire radial extension of NGC 1904, we then needed to
combine the MUSE RVs thus measured, with the catalog obtained by 
\citet{ferraro+18b} from FLAMES and KMOS observations. We assumed as ``reference'' RV values those
measured from the FLAMES dataset, because the spectral resolution of
this instrument is the highest and thus provides the most accurate and
reliable measures.  By using the stars in common between the MUSE/WFM
and the FLAMES samples, we found just a very small average offset
($-1.0$ km s$^{-1}$), which was then applied to the WFM sample for
realigning it with the reference catalog. A residual $-0.5$ km
s$^{-1}$ offset was then identified between the RV values measured for
the stars in common between the NFM and the (realigned) WFM data
sets. After applying this small offset to the NFM data, we finally
obtained three homogeneous samples of RVs.  To create the final
catalog, where every star is assigned with a single RV value, we
adopted the following criteria aimed at adopting the best available
measure in each case: for the stars in common between FLAMES and
MUSE/WFM, we selected the RV measurements obtained with FLAMES, for
the stars in common between KMOS and MUSE, we used the MUSE measures,
and in case of common targets between MUSE/WFM and MUSE/NFM, we used
the NFM values.

\begin{figure}[ht!]
\centering \includegraphics[width=19.2cm, height=8.2cm]{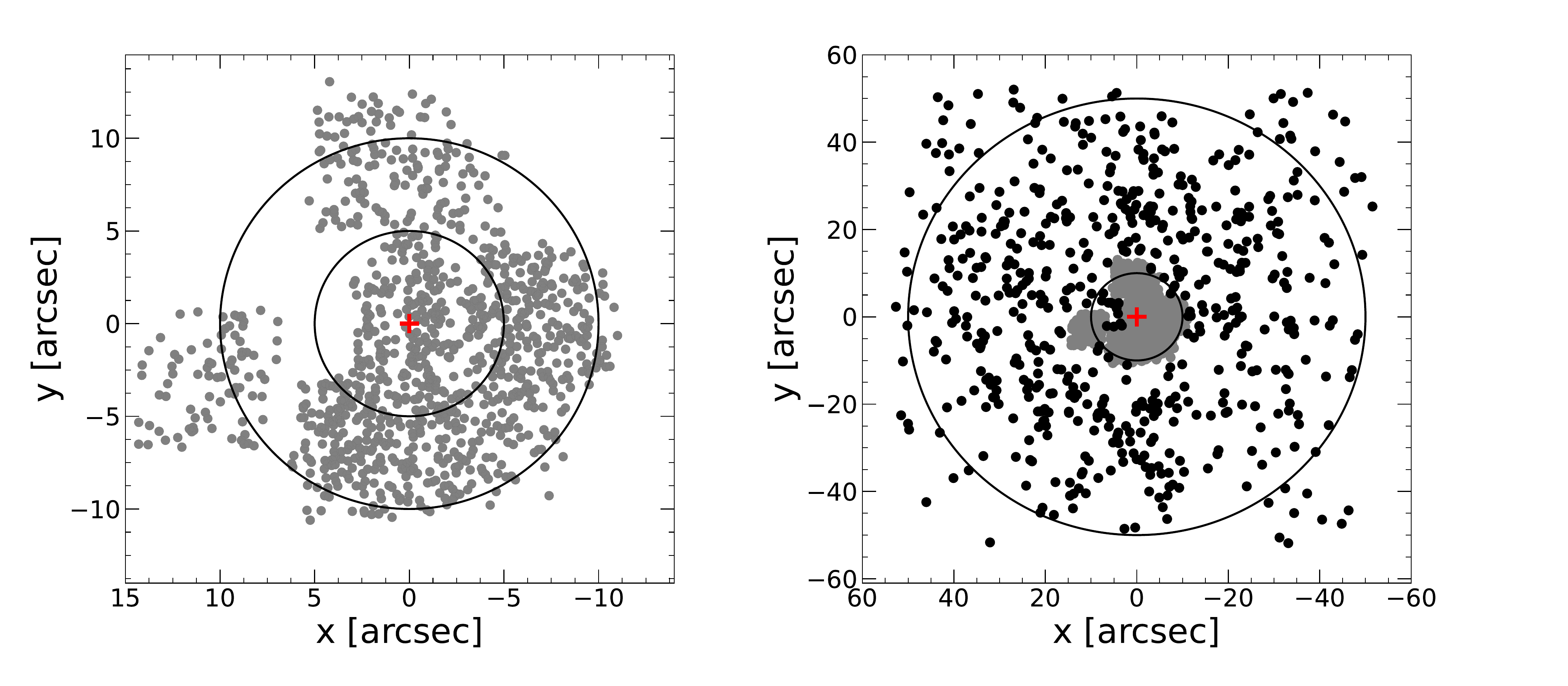}
\caption{{\it Left panel:} map on the plane of the sky, with respect
  to the adopted cluster center (red cross), for all the stars with RVs measured from
  MUSE/NFM data. The two circles mark distances of $5\arcsec$ and
  $10\arcsec$ from the center. {\it Right panel:} the same, but for
  the stars with RVs obtained from the MUSE/WFM observations (black
  dots). The MUSE/NFM sample is also shown in gray and the two circles mark distances of $10\arcsec$ and
  $50\arcsec$ from the center (red cross).}
\label{fig:map}
\end{figure}

\begin{figure}[ht!]
\centering
\includegraphics[width=17cm, height=10cm]{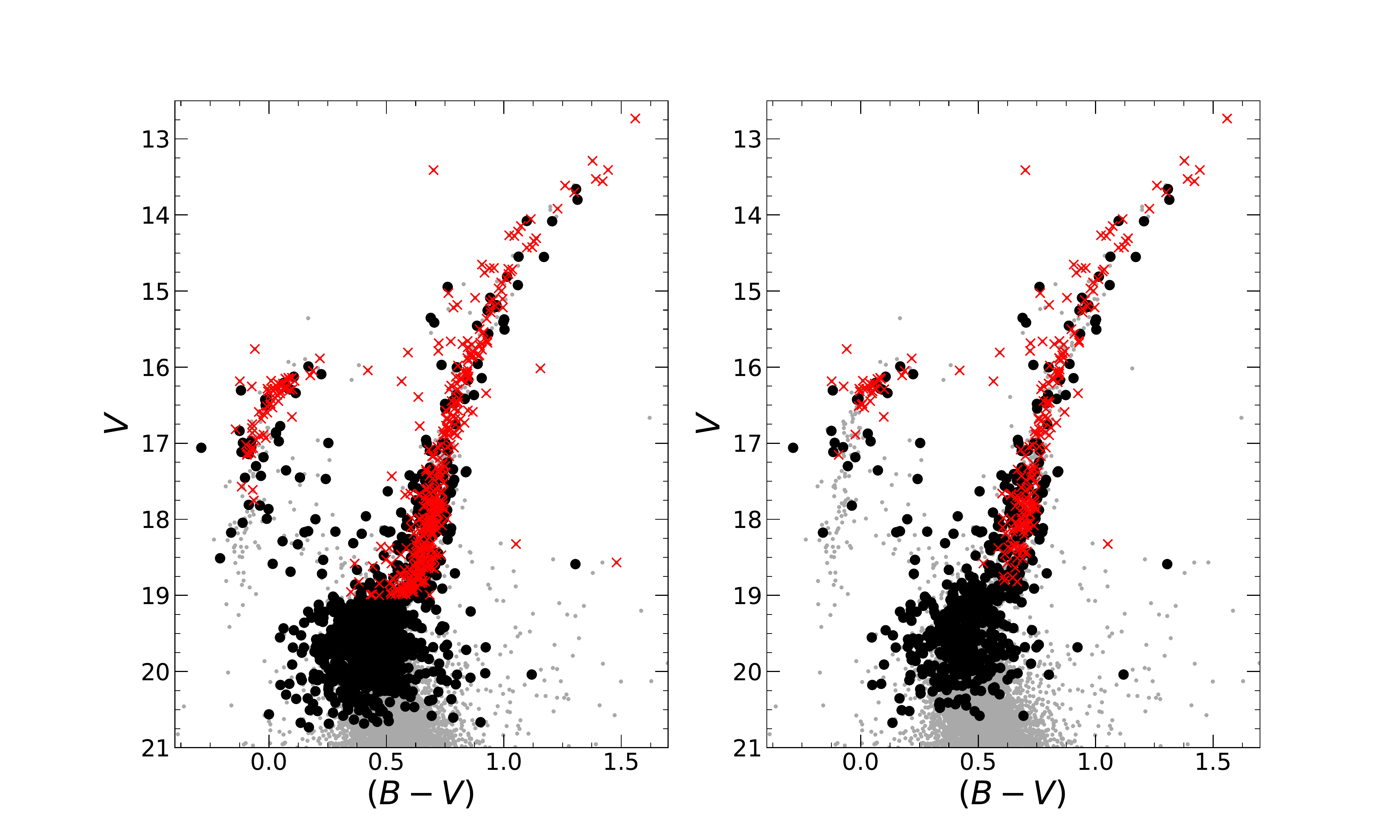}
\centering
\caption{{\it Left panel:} CMD of NGC 1904 obtained from the
  photometric catalog discussed in \cite{lanzoni+07} (gray dots),
  with the stars having RVs measured from the MUSE observations marked
  as black circles (NFM sample) and red crosses (WFM dataset).  {\it
    Right panel:} the same, but only for the stars used to determine
  the internal kinematics of NGC 1904, i.e., stars with RV uncertainty
  $\epsilon_{RV}<10$ km s$^{-1}$, S/N$>10$ and contamination ratio $C\le 0.05$.}
\label{fig:cmd}
\end{figure}

The final catalog lists 1726 individual stars with measured RV, each
dataset contributing as follows:
\begin{itemize}
\item[(1)] The MUSE/NFM catalog consists of 946 measures for sources
  located between $0.3\arcsec$ and $15.7\arcsec$ from the cluster center
  (Figure \ref{fig:map}, left panel). They sample the magnitude range
  $13.5 < V < 21$, and their position in the $(V, B-V)$ CMD is shown
  in left panel of Figure \ref{fig:cmd} (black dots).
\item[(2)] The MUSE/WFM dataset consists of 587 stars located between
  $2.3\arcsec$ and $66.5\arcsec$ from the cluster center (black dots
  in the right panel of Figure \ref{fig:map}) and with magnitudes
  $12<V<19$ (red crosses in the left panel of Figure \ref{fig:cmd}).
\item[(3)] The FLAMES/KMOS sample consists of 193 stars, sampling the
  radial region beyond $\sim 20\arcsec$, out to $774\arcsec$ from the
  cluster center \citep[see][]{ferraro+18b}.
\end{itemize}

\subsection{Contamination estimate of the MUSE targets}
\label{sec:contamination}
Although in NGC 1904 we do not expect significant contamination from the 
foreground Galactic field sources, the MUSE observations presented in this paper sample
the core of the cluster, where stellar crowding is critical. 
Hence, the RV value measured for a
given star might be affected by the presence of bright neighboring
sources. In turn, this can impact the final results, in terms of
velocity dispersion and systemic rotation of the cluster.
To quantify this effect and select the spectra contributed by the
light of individual stars only, we implemented a procedure aimed at
estimating the level of contamination suffered by each MUSE target.

We considered all the sources listed in the HST catalogs used for the
photometric analysis and we modelled each of them with the adopted MUSE PSF function
(see Section \ref{sec:selection}). For every MUSE target,
this allowed us to estimate the amount of light in the central spaxel
contributed by the star under analysis and the surrounding
objects. Using the entire HST catalogs (instead of only the MUSE
targets with measured RV) guarantees that we are taking into account a
complete list of sources, including stars that are not identified
(because they are too faint) or are blended in the MUSE observations,
as well as stars that are located just beyond the edges of the MUSE
fields. From this analysis we then quantify the level of contamination
of each MUSE target through the contamination parameter ($C$) defined
as the ratio between the fraction of light contributed by the first
contaminant and that of the target under analysis, where the first
contaminant is the neighbouring source providing the largest
contribution to the central spaxel light after the target itself.

\begin{figure}[ht!]
\centering
\includegraphics[width=17.5cm, height=5.3cm]{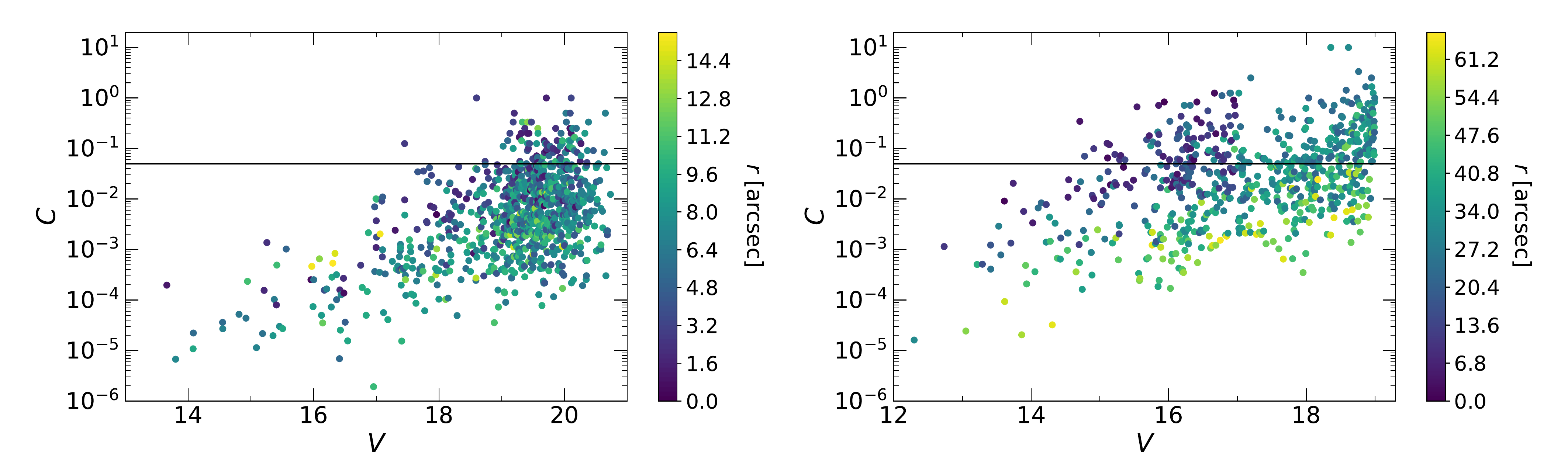}
\caption{Contamination parameter ($C$, see Section
  \ref{sec:contamination}) as a function of magnitude, for the
  MUSE/NFM targets (left panel) and the MUSE/WFM sources (right
  panel). The color scale indicates the distance of the targets from
  the cluster center (see color bars). The black line flags the
  adopted contamination threshold: only targets with $C\le 0.05$ have
  been used to determine the internal cluster kinematics.}
\label{fig:contam}
\end{figure}

Figure \ref{fig:contam} shows the parameter $C$ estimated for all the
MUSE targets as a function of the magnitude, with the color code
illustrating the dependence on the distance from the cluster center.
As expected, the most contaminated stars (large values of $C$) are the
faint ones. Moreover, the contamination from neighbouring stars is
more severe for the observations acquired at lower spatial resolution
(WFM, right panel), with respect to those performed with AO correction
(NFM, left panel). The figure also illustrates that, to discriminate
isolated from contaminated targets, it is more efficient to select a
threshold in $C$, instead of a cut in magnitude; indeed, this allows
us to keep in the sample faint but isolated stars, as well as to
reject targets that, in spite of being relatively bright, are
contaminated by neighbors.  For the following analysis, we made the
conservative choice of keeping in the sample only the MUSE targets
with contamination parameter smaller than 5\% ($C\le 0.05$), for which
we estimate that neighboring stars have a negligible impact on the
measured RV. The final sample includes 857 stars in the NFM dataset,
and 351 stars observed with MUSE/WFM.

\section{Results}
\label{sec:result}
To properly determine the internal kinematics (systemic velocity,
velocity dispersion profile and rotation curve) of NGC 1904 from the
available dataset, we further selected the sample by considering only
stars with RV uncertainty smaller than 10 km s$^{-1}$ and S/N higher than 10. This brought
the final sample to a total of 1078 RV measures, for individual stars
located in a large range of radial distances, from $0.3\arcsec$ out to
$\sim 774\arcsec$ (about 14 three-dimensional half-mass radii, assuming the half-mass radius of the system $r_h=56.7\arcsec$ presented in \citealt{miocchi+13}) from the cluster center. The CMD distribution of the
selected MUSE target is shown in the right panel of Figure
\ref{fig:cmd}.

\subsection{Systemic velocity}
\label{sec:vsys}
\begin{figure}[ht!]
\includegraphics[width=17cm, height=6cm]{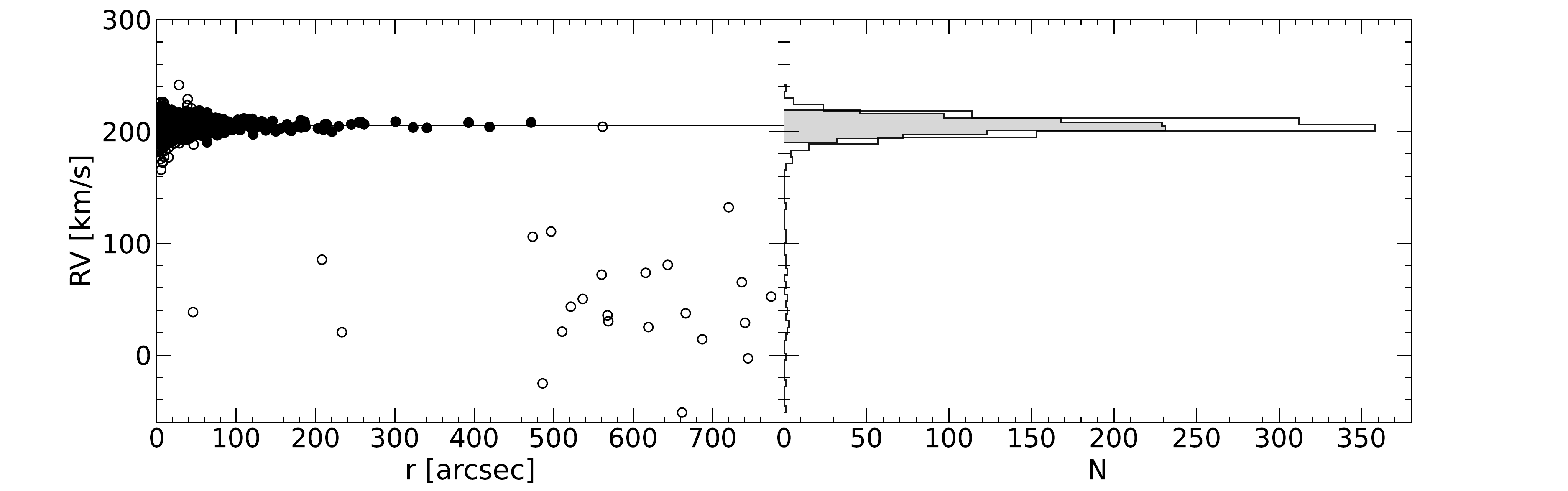}
\centering
\caption{{\it Left panel:} Radial velocities of the final sample of
  (1078) selected stars as a function of their distance from the
  cluster center. The solid circles flag those used to determine the
  cluster systemic velocity (black solid line), while the empty
  circles mark those rejected by the 3$\sigma$-clipping algorithm.
  The number distribution of the targets used for the systemic
  velocity is plotted as a filled gray histogram and compared to that
  of the entire sample (empty histogram) in the right panel.}
\label{fig:rv_dist}
\end{figure}

Figure \ref{fig:rv_dist} shows the distribution of the 1078 RVs
measured in NGC 1904 as a function of the distance from the cluster
center.  The population of cluster members is clearly distinguishable
as a narrow, strongly peaked component, which dominates the sample at
RV $\sim200$ km s$^{-1}$. To carefully determine the value of the
cluster systemic velocity ($V_{\rm sys}$), we rejected obvious
outliers, selecting only the targets with RV in the range 190 km
s$^{-1} <$ RV $<$ 220 km s$^{-1}$.  Under the assumption that the RV
distribution is Gaussian, we used a Maximum-Likelihood approach (e.g.,
\citealt{Walker_2006}) to estimate the cluster systemic velocity and its
uncertainty. By applying the selection criteria described above and a
$3\sigma$-clipping procedure, a sample of 998 RV measures has been
used for the estimate of $V_{\rm sys}$. These are marked as black
solid circles in the left panel of Figure \ref{fig:rv_dist} and their
distribution is shown in grey in the right panel of the same figure.
The resulting value is $V_{\rm sys} \ 205.4 \pm 0.2$ km s$^{-1}$, in good agreement with previous determinations, e.g., $205.8 \pm 0.4$ km s$^{-1}$ \citep{harris96}, $205.78 \pm 0.54$ km s$^{-1}$ \citep{dorazi+15}, $205.4 \pm 0.6$ km s$^{-1}$ \citep{ferraro+18b} and $205.6 \pm 0.2$ km s$^{-1}$ \citep{Baumgardt+18}.
Different,
but still reasonable, assumptions about the cuts in RV, error and
contamination parameter produce no significant variations in this
result.  We also verified that using only the MUSE/NFM and MUSE/WFM
catalogs individually provides perfectly consistent values of $V_{\rm
  sys}$, as illustrated in Table \ref{tab:vsys}. 
  In the following, we will use $V_r$ = RV $-$ $V_{\rm sys}$ to
indicate the RV values referred to the cluster systemic velocity.

\begin{deluxetable*}{lRRRRRR}
\tablecaption{Systemic velocity of NGC 1904}
\tablewidth{0pt}
\tablehead{
\colhead{ Catalog }  & \colhead{$r_{min}$}  &  \colhead{$r_{max}$}  & 
\colhead{$N$} & \colhead{$N_{V}$} & \colhead{ $V_{sys}$ }\\  
\colhead{  } & \colhead{[arcsec]}  &  \colhead{[arcsec]}  &
\colhead{      } & \colhead{      } & \colhead{ km s$^{-1}$ } }
\startdata
MUSE/NFM  & 0.3 & 15.7 & 599 & 553 &  205.2 $\pm$ 0.3\\
MUSE/WFM & 4.2 & 64.1 & 286  & 276 & 205.7 $\pm$ 0.4 \\
MUSE + FLAMES/KMOS & 0.3 & 773.7 & 1078 & 998  & 205.4 $\pm$ 0.2  \\
\enddata
\tablecomments{For the catalog including only MUSE/NFM targets, that
  of MUSE/WFM stars, and the global catalog obtained after the
  combination of the previous two with the catalog discussed in
  \citet{ferraro+18b}, the table lists the minimum and maximum
  distances from the cluster center sampled by the observed stars
  ($r_{\rm min}$ and $r_{\rm max}$, respectively), total number of
  stars after all selections (i.e., RV uncertainty $\epsilon_{RV}<10$ km s$^{-1}$, S/N $>10$ and $C\le0.05$), the number of stars used for the
  determination of the systemic velocity ($N_V$), and the resulting
  value of $V_{\rm sys}$ with its 1$\sigma$ uncertainty.}
\label{tab:vsys}
\end{deluxetable*}

\subsection{Second velocity moment profile}
\label{sec:2ndv}
The dispersion of the measured RVs about the cluster systemic
velocity, determined at different radial distances from the center,
provides the second velocity moment profile of the system,
$\sigma_{II}(r)$. Under the assumption of no relevant rotation, this
is a good approximation of the projected velocity dispersion profile,
$\sigma^2_P(r)$. In fact:
\begin{equation}
  \sigma^2_P(r) = \sigma^2_{II}(r) - A^2_{\rm rot}(r),
\end{equation}
where $A_{\rm rot}$ is the rotation curve amplitude. Indeed,
preliminary evidence of rotation was recently detected in the external
region of NGC 1904 \citep{ferraro+18b}, and it is confirmed in the
present study, as discussed in the next section. 
Hence, the correct determination of the velocity dispersion profile
of this system requires the evaluation of the strength of rotation 
and its radial variation.
However, the second velocity moment profile is still worth to be
determined since it offers the opportunity of a first-order comparison
with previous results in literature that ignored the effects of the
cluster rotation.

As usual, the radial profile has been obtained by splitting the
surveyed area into a set of concentric annuli, at increasing distance
from the cluster center, chosen as a compromise between a good radial
sampling and a statistically significant number of stars (at least
20 - 30).  In each radial bin, obvious outliers (like field stars,
having RVs in clear disagreement with the cluster distribution in that
radial interval) have been excluded from the analysis and a
3$\sigma$-clipping algorithm about the cluster systemic velocity has
been applied to further clean the sample. Then, $\sigma_{II}(r)$ has
been computed from the dispersion of the values of $V_r$
measured for all the remaining stars in each annulus, by following the
Maximum-Likelihood method described in \citeauthor{walker+06}
(\citeyear{walker+06}, see also \citealt{Martin_2007};
\citealt{10.1111/j.1365-2966.2009.14864.x}). The error is estimated 
following the procedure outlined in \citet{1993ASPC...50..357P}.

\begin{figure}[ht!]
\centering
\includegraphics[width=19.5cm, height=10cm]{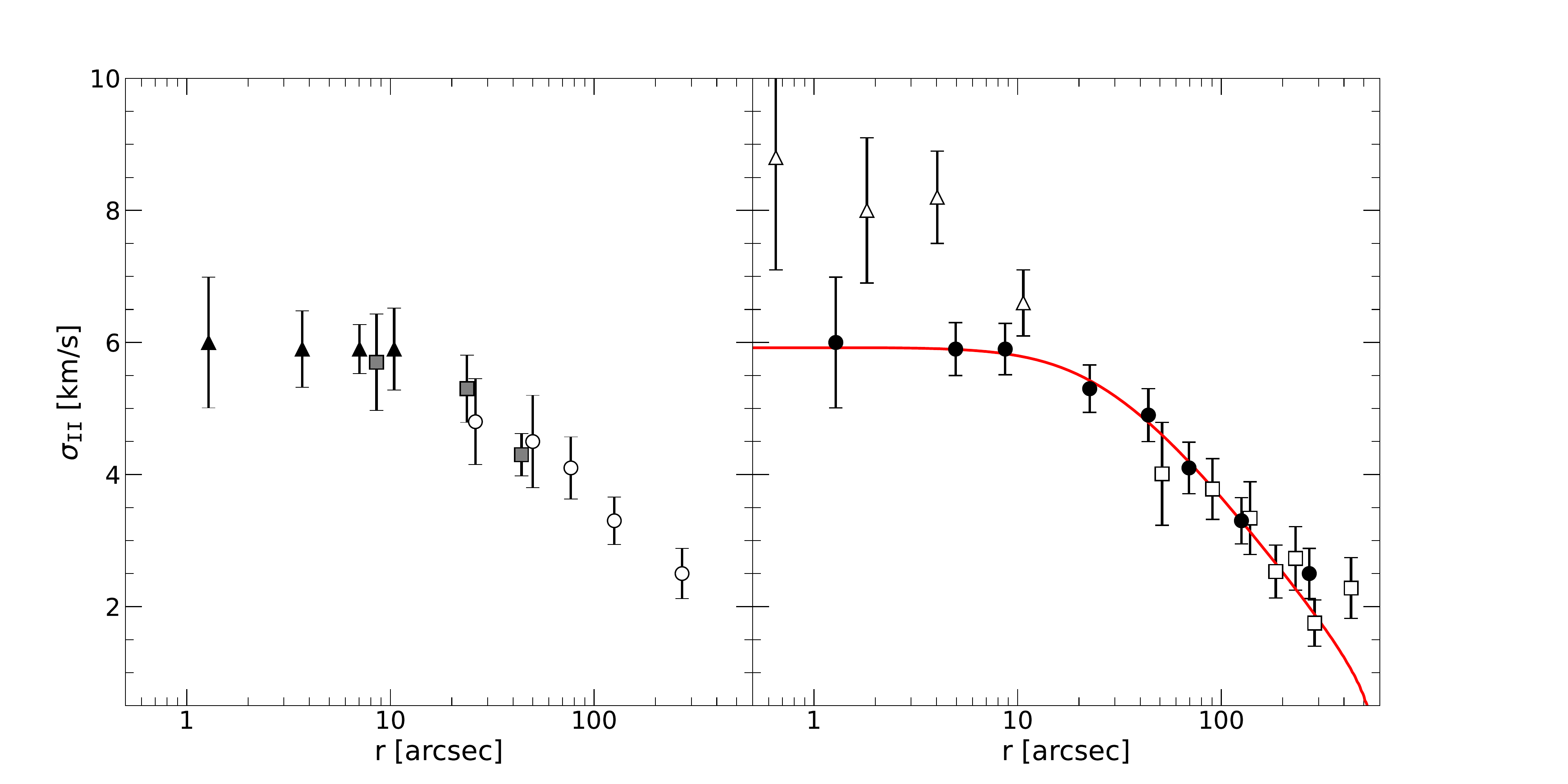}
\centering
\caption{{\it Left panel:} Second velocity moment profiles obtained
  from the three considered catalogs separately: MUSE/NFM data (black triangles),
  MUSE/WFM data (gray squares) and the catalog
  presented in \citet[][empty circles]{ferraro+18b}.
  {\it Right panel:} Second velocity moment profile of NGC 1904 obtained from
  the individual RVs of all the stars included in the combined
  MUSE/NFM, MUSE/WFM and FLAMES dataset (solid circles).
  The empty triangles represent the profile derived by \citet{lutz+13}, from the line broadening of integrated-light spectra in the inner regions, while the empty squares have been obtained from individual RVs by \citet{scarpa+11} in the outskirts. The red solid line shows the King model that best fits the star density profile of NGC 1904, as obtained in \citet{miocchi+13}.}
\label{fig:2ndv_combi}
\end{figure}

For the sake of illustration, the left panel of Figure
\ref{fig:2ndv_combi} shows the second velocity moment profiles obtained
from the three considered catalogs, separately: MUSE/NFM and MUSE/WFM
discussed in this study (black triangles and gray squares,
respectively) and the FLAMES-KMOS catalog of \citet[][empty circles]{ferraro+18b}. Table \ref{tab:2ndv_01} lists the values
obtained. The figure illustrates how the three catalogs provide
complementary coverages of different radial portions of the cluster,
and that the resulting kinematic profiles are in very good agreement
each other. The final second velocity moment profile of NGC 1904, as
obtained from the combined catalog, is shown in the right panel of Figure \ref{fig:2ndv_combi} (solid circles) and listed in Table \ref{tab:2ndv_01}. Its stays flat in the central regions ($r \lesssim 10\arcsec$) with a
central value of $\sim$ 6.0 km s$^{-1}$, then decreases in the outer
regions, as usually observed.
 
In the right panel of Figure \ref{fig:2ndv_combi}, this profile is compared with that published
in \citet[][empty triangles]{lutz+13}, which was obtained by measuring
the line broadening of integrated-light spectra at $r<10$\arcsec, and with that obtained in \citet[][empty squares]{scarpa+11} from individual RVs in the outer cluster region. As can
be seen, while the profiles agree quite well at large distances from
the center, the value measured in the innermost core is significantly
larger than that presented here, and it was interpreted as the
signature of a $\sim 3000 M_\odot$ IMBH. This discrepancy 
is qualitatively similar, although significantly less pronounced, to that found
in the case of NGC 6388, where it was ascribed to the presence of two
bright central stars with opposite RVs provoking a spurious broadening
of the (integrated-light) spectral lines and, in turn, an
over-estimate of the velocity dispersion in the cluster innermost
region \citep[see][]{lanzoni+13}. In the innermost $10\arcsec$ of NGC
1904 we measured the RV for almost 530 stars. Hence, we conclude that
the result presented in the right panel of Figure \ref{fig:2ndv_combi} is solid. It indicates a
central value of the second velocity moment $\sim 25\%$ smaller than
that quoted in \citet[][$\sigma_0=6 \ \rm km \ s^{-1} \ instead \ 8 \ km \ s^{-1}$]{lutz+13},
and it excludes the presence of a central IMBH, at least an IMBH
massive enough to produce a detectable perturbation in the second
velocity moment profile. We also note that, in spite of having adopted
the results of \citet{lutz+13}, the best-fitting {\it N}-body model to the
surface density and line-of-sight velocity dispersion profiles of NGC
1904 determined by \citet{Baumgardt+18} suggests a central value of $\sim
6.5$ km s$^{-1}$ (see their Figure E1), which is in good agreement
with our analysis.

\begin{deluxetable*}{RRRRCC}
\tablecaption{Second velocity moment profiles obtained from the
  MUSE/NFM and MUSE/WFM datasets, separately, and using the
  combined catalog of MUSE/NFM, MUSE/WFM and FLAMES data.}
\tablewidth{0pt}
\tablehead{
\colhead{ $r_i$ } & \colhead{ $r_e$ }  & \colhead{$r_m$}  &
\colhead{$N$} & \colhead{$\sigma_{II}$} & \colhead{$\epsilon_{\sigma_{II}}$} \\  
\colhead{ [arcsec] } & \colhead{ [arcsec] }  & \colhead{[arcsec]}  &  \colhead{ } &\colhead{km s$^{-1}$ }  &
\colhead{ km s$^{-1}$} 
}
\startdata
\hline
\multicolumn{6}{c}{MUSE/NFM}\\
\hline
0.01  &  2.00   & 1.28  & 31  &  6.00 & 0.99\\
2.00  &   5.00  &  3.69 & 121 &  5.90 & 0.58  \\
5.00  &   9.00  &  7.05 & 294 &  5.90 & 0.37  \\
9.00  &  15.00  & 10.43 & 125 &  5.90 & 0.62 \\
\hline
\multicolumn{6}{c}{MUSE/WFM}\\
\hline
 0.01  & 15.00 &  8.54  &  35  &  5.70 & 0.73   \\
15.00 &  30.00 &  23.74 &  88  &  5.30 & 0.51   \\
30.00  & 66.00 &  44.00 & 187  &  4.30 & 0.32   \\
\hline
\multicolumn{6}{c}{Combined Catalog}\\
\hline
0.01  &  2.00   &  1.28  &  31 &  6.00 & 0.99  \\
2.00  &   7.00  &  4.96  & 269 &  5.90 & 0.40  \\
7.00  &  11.00  &  8.69  & 259 &  5.90 & 0.39  \\
11.00 &  35.00  &  22.58 & 203 &  5.30 & 0.36  \\
35.00 &  55.00  & 43.84  & 138 &  4.90 & 0.40  \\
55.00 & 100.00  & 69.37  & 73  &  4.10 & 0.39  \\
100.00 & 170.00 & 125.57 & 43  &  3.30 & 0.35  \\
170.00 & 600.00 & 270.13 & 23  &  2.50 & 0.38  \\
\enddata
\tablecomments{The first three columns list the internal, external,
  and mean radii of each adopted radial bin ($r_i$, $r_e$ and $r_m$,
  respectively), with the mean radius computed as the average distance
  from the center of all the stars in the bin ($N$, fourth
  column). The last two columns list the second velocity moment and
  its uncertainty in each bin, respectively.}
\label{tab:2ndv_01}
\end{deluxetable*}

\newpage
\subsection{Systemic rotation}
\label{sec_vrot}
A significant signature of rotation has been found in the outer
regions of NGC 1904, from the analysis of individual RVs by
\citet{ferraro+18b}. This is consistent with the findings presented in
\citet{scarpa+11}, who measured a cluster rotational velocity of 1.1
km s$^{-1}$ within $3\arcmin$ from the center, with respect to an axis
with position angle PA$=85\arcdeg$ (as measured from North toward East),
from a total sample of $\sim 150$ RVs.  Signatures of rotation in the
plane of the sky, as determined from the analysis of proper motions,
appears more controversial. A signal of rotation has been presented in
the recent compilation of \citet{vasiliev+21}, while
\citet{sollima+19} classified the cluster rotation as uncertain,
because, although a signal was detected at more than 3$\sigma$ significance 
level, the GC failed one of the tests performed to check for spurious 
rotation effects. Thanks to the large sample of MUSE
individual RVs presented in this paper, we are now able to explore the
rotational properties of NGC 1904 over its entire radial extension. On
the other hand the analysis of the Gaia EDR3 offers the opportunity to
study the kinematics also in the plane of the sky, thus providing a
comprehensive analysis of the cluster rotation.

\begin{figure}[ht!]
\includegraphics[width=19cm, height=13.0cm]{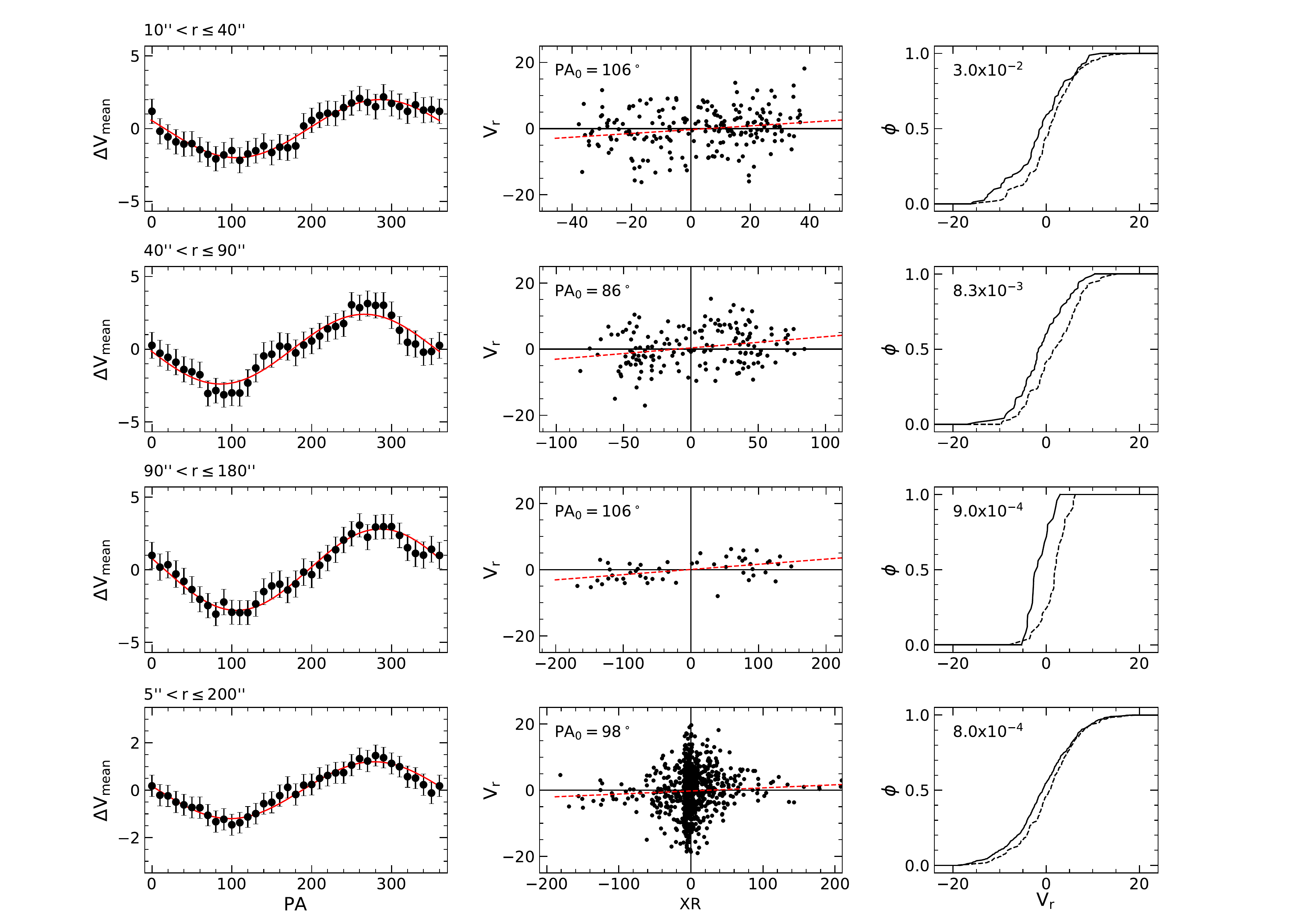}
\centering
\caption{Diagnostic diagrams of the rotation signature detected in three concentric annuli at different distances from the cluster center (three top panels; see labels in their top-left corner), and considering the entire sample with $r>5\arcsec$ (bottom panels). For each bin the {\it left panels } show the difference between the mean RV on each side of a line passing through the center with a given position angle (PA),
as a function of PA itself. The continuous line is the sine function that best fits the observed patterns. 
The {\it central panels} show the distribution of the radial velocities $V_r$ as a function
of the projected distances from the rotation axis (XR) in arcseconds. The position angle of the rotation axis (PA$_0$) is labeled in each panel. The red dashed lines are the least square fits to the data. 
The {\it right panels} show the cumulative RV distributions for the sample of stars
with XR$<0$ (solid line) and for that with XR$>0$ (dotted line). The Kolmogorov-Smirnov probability that the two samples on each side of the rotation axis are drawn from the same parent distribution is also labelled.}
\label{fig:fig_vrot_bin_glob}
\end{figure}

For the line-of-sight direction, we adopted the same approach followed
in \citet{ferraro+18b, lanzoni+18a, lanzoni+18b} and described, e.g.,
in \citet[][see also \citealp{lanzoni+13}]{bellazzini+12}.  The method
consists in splitting the RV dataset in two sub-samples by means of a separation line passing through the cluster center and with position
angle (PA) varying between $0\arcdeg$ (North direction) and
$180\arcdeg$ (South direction), in steps of $10\arcdeg$, with
PA$=90\arcdeg$ corresponding to the East. For each value of PA, the
difference between the mean velocities of two groups of stars ($\Delta
V_{\rm mean}$) is measured and recorded.  By construction, the method
requires a sample of RVs symmetrically distributed on the plane of the
sky, while it cannot be applied if large areas remain unsampled by the
available dataset. The presence of rotation along the line-of-sight
is expected to produce the following set of observable properties:
\begin{itemize}
\item[(1)] By progressively rotating the separating line from North to
  South, $\Delta V_{\rm mean}$ is expected to draw a coherent
  sinusoidal dependence on PA.  The maximum/minimum of this curve
  provides the position angle of the rotation axis (PA$_0$), since it
  corresponds to the strongest separation in an approaching and a
  receding sub-samples.  The absolute value of the maximum/minimum of
  this curve corresponds to twice the rotation amplitude ($A_{\rm
    rot}$).
\item[(2)] The projected spatial distribution is expected to be 
 flattened in the direction of the rotation axis.
\item[(3)] The sub-samples of stars on each side of the rotation axis
  are expected to
  show, not only different mean velocities (as quantified by $\Delta
  V_{\rm mean}(\rm PA_0)$), but also different cumulative $V_r$ distributions.
  To quantify the statistical significance of such
  difference we used three estimators: the probability that the RV
  distributions of the two sub-samples are extracted from the same
  parent family is evaluated by means of a Kolmogorov-Smirnov test,
  while the statistical significance of the difference between the two
  sample means is estimated with both the Student's t-test and a
  Maximum-Likelihood approach.
\end{itemize}
We applied this procedure to our RVs sample, in three concentric annuli
around the cluster center with sufficiently symmetric spatial sampling
(we thus excluded the innermost $10\arcsec$: see the left panel of
Figure \ref{fig:map}).
The results are plotted in Figure \ref{fig:fig_vrot_bin_glob}. In all
the considered annuli, we find well-defined sinusoidal behaviors of
$\Delta V_{\rm mean}$ as a function of PA (left-hand panels),
asymmetric distributions of $V_r$ as a function of the projected
distance from the rotation axis XR (central panels), and
well-separated cumulative $V_r$ distributions for the two samples on
either side of the rotation axis (right-hand panels). The reliability
of these systemic rotation signatures is also confirmed by the values
of the Kolmogorov-Smirnov and t-Student probabilities and by the
significance level of different sample means obtained from the
Maximum-Likelihood approach (see the three last columns in Table
\ref{tab_vrot_anelli_glob}), thus providing a solid confirmation of a
coherent global rotation of the system.
Averaging values of PA$_0$ weighted by the number of stars in each
radial bin, we obtained PA$_0=98\arcdeg$, which is adopted as
global position angle for the cluster rotation axis.  This locates the
rotation axis essentially aligned to the East-West direction, dividing
the observed dataset into a northern sub-sample with positive mean $V_r$,
and a southern approaching sub-sample (with negative mean $V_r$).  By
fixing PA$_0$ to this value, we applied the described procedure to all
stars measured at $r>5\arcsec$ (for the reason discussed in the next
paragraph), finally obtaining the diagnostic plots shown in the three bottom panels of Figure \ref{fig:fig_vrot_bin_glob} and the results listed in the bottom line of Table \ref{tab_vrot_anelli_glob}.

\begin{deluxetable*}{rrrrccccc}[ht!]
\tablecaption{Rotation signatures detected in circular annuli around
  the cluster center and global rotation found in NGC 1904 using the entire sample with $r>5\arcsec$ (bottom line). \label{tab_vrot_anelli_glob}}
\tablewidth{0pt}
\tablehead{
\colhead{$r_i$ } & \colhead{ $r_e$ } & \colhead{ $r_m$ } & \colhead{ $N$}  &  \colhead{PA$_0$}  &
\colhead{ $A_{\rm rot}$} & \colhead{ $P_{\rm KS}$}  & \colhead{ $P_{\rm Stud}$} & \colhead{n-$\sigma_{\rm ML}$}
}
\startdata
   10  &   40  &  27.3 &  206 &  106  &   1.0 & $3.0\times 10^{-2}$ &  $>95.0$  & 2.2  \\ 
   40  &   90  &  54.7 &  165 &   86  &   1.2 & $8.3\times 10^{-3}$ &  $>99.8$  & 3.1  \\ 
   90  &  180  & 122.2 &   51 &  106  &   1.4 & $9.0\times 10^{-4}$ &  $>99.8$  & 4.0  \\
   \hline
   5  &  800  &  34.6 &  881 &   98  &   0.6 & $8.0\times 10^{-4}$ &  $>99.0$  & 2.9  \\
\enddata
\tablecomments{For each radial bin the table lists: the inner and outer
  radius ($r_i$ and $r_e$) in arcseconds, the mean radius and the
  number of stars in the bin ($r_m$ and $N$, respectively), the
  position angle of the rotation axis (PA$_0$), the rotation amplitude
  (A$_{\rm rot}$), the Kolmogorov-Smirnov probability that the two
  samples on each side of the rotation axis are drawn from the same
  parent distribution ($P_{\rm KS}$), the t-Student probability that
  the two RV samples have different means ($P_{\rm Stud}$), and the
  significance level (in units of n-$\sigma$) that the two means are
  different following a Maximum-Likelihood approach (n-$\sigma_{\rm
    ML}$).}
\end{deluxetable*}

\begin{figure}[ht!]
\includegraphics[width=10.2cm, height=9.5cm]{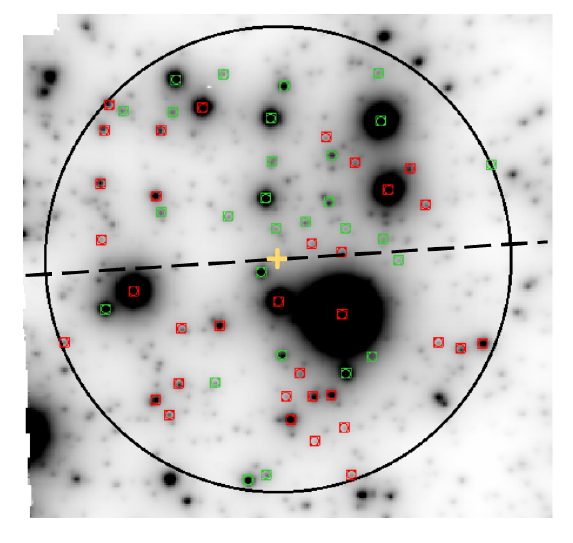}
\centering
\caption{Zoomed portion of the reconstructed MUSE/NFM image in the
  innermost region of NGC 1904.  The black circle is center on the
  cluster center (yellow cross) and has a radius of $3\arcsec$. North
  is up, East is left. Red symbols mark stars with RV larger than the
  cluster systemic velocity (receding stars), while green symbols mark
  the stars approaching the observer. The orientation (PA$_0 = 93 \arcdeg$)
  of the axis that maximizes the difference between the mean RVs of the 
  two sub-samples is also shown as a dashed line.
\label{fig:fig_inner3}}
\end{figure}

{\it The peculiar behaviour of the innermost core -} In performing the
analysis of the rotation curve, we discovered a peculiar feature that
needs to be further investigated.  Since the sampling of the innermost
$3\arcsec$ is sufficiently symmetric (see the left panel of Figure
\ref{fig:map}), we applied the described procedure also in this
region, leaving the position angle free to vary between $0\arcdeg$ and
$180\arcdeg$. We found evidence of a rotation signal for PA$_0=
93\arcdeg$, but the rotation pattern is just opposite to that found
for $r>10\arcsec$: the northern sub-sample has negative mean $V_r$ (it is
approaching), while the the stars observed in the southern hemisphere
are preferentially receding from the observer (positive mean $V_r$).
Admittedly the detection of this effect is based on a just limited
number of stars ($\sim 60$) and the statistical significance is not
high (e.g., t-Student probability that the two RV samples have
different means is $P_{\rm Stud}\sim 90\%$, only).  
Figure \ref{fig:fig_inner3}
shows a MUSE/NFM image zoomed in the innermost $3\arcsec$ from the
center with the considered stars flagged with two different colors: in
red those receding from the observer, in green the approaching
ones. The difference in the mean RV of the two samples is
apparent also by eye.  
If confirmed, this could represent the detection of a kinematically decoupled core. 
A similar kinematic feature has been detected so far only in M15 
\citep[]{van_den_Bosch_2006, bianchini+13, kamann+18, Usher_2021}. 
It has been suggested that stellar interactions with a massive 
binary black holes \citep{mapelli2005} may produce this feature
but the viability of this scenario in M15 is unclear and further 
investigations of other dynamical scenarios and the possible role 
of core collapse are necessary.
In any case, the statistics is
admittedly too poor to draw any firm conclusion. We therefore avoid any
further discussion of this feature and we exclude the stars located at
$r<3\arcsec$ from the following analysis.

\subsection{Line-of-sight rotation curve}
\label{sec_kin}
\begin{figure}[ht!]
\includegraphics[width=14.8cm, height=8.3cm]{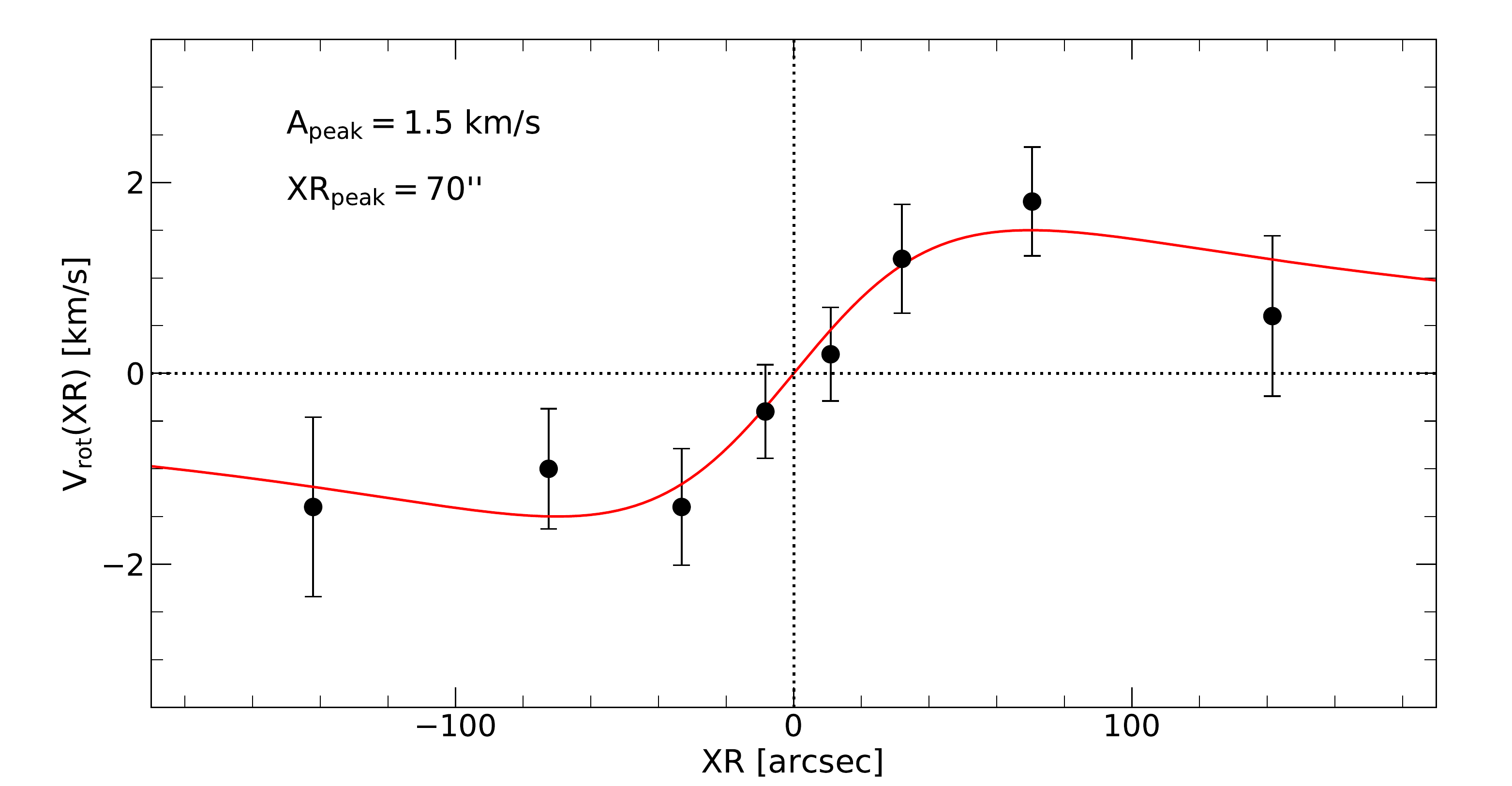}
\centering
\caption{Rotation curve of NGC 1904. The black circles mark the
  stellar mean velocity as a function of the projected distance on
  either side of the rotation axis (XR) for the intervals listed in
  Table \ref{tab_rotcurve}. The red line, which well reproduces the
  observed curve, is the model shown in equation (\ref{eq_curve}),
  with $A_{\rm peak}=1.5$ km s$^{-1}$ and ${\rm
    XR_{peak}}=70\arcsec$.}
\label{fig:fig_rotcurve}
\end{figure}

We built the rotation curve of NGC 1904 by splitting the RV sample in
four intervals sampling increasing values of XR on both sides of the
rotation axis.  Following \citet{walker+06, sollima+09, lanzoni+18a},
we used a Maximum-Likelihood method to determine the mean velocity of
all the stars belonging to each XR bin.  The errors have been
estimated following \citet{pryor+93}.  The resulting rotation curve is
shown in Figure \ref{fig:fig_rotcurve} and presented in Table
\ref{tab_rotcurve}. It clearly shows the expected shape, with a steep
increasing trend in the innermost regions, and a maximum value
followed by a soft decreasing behavior outward.  The observed trend is
well reproduced by an analytic function \citep{lyndenbell67}
appropriate for cylindrical rotation:
\begin{equation}
  V_{\rm rot} = \frac{2 A_{\rm peak}}{\rm XR_{peak}} ~~ \frac{\rm
    XR}{1 + ({\rm XR}/{\rm XR_{peak}})^2,}
\label{eq_curve}  
\end{equation}
where $A_{\rm peak}$ and XR$_{\rm peak}$ are the maximum rotation
amplitude and its distance from the rotation axis, respectively.  The
red solid line in Figure \ref{fig:fig_rotcurve} shows the model
computed for a maximum amplitude of $\sim 1.5$ km s$^{-1}$ at XR$_{\rm 
peak}\sim 70\arcsec$ from the rotation axis, approximately corresponding to the three-dimensional half-mass radius (or nearly 2 projected half-mass radii, adopting the value $R_{\rm h}=41.7$\arcsec \ quoted in \citealt{miocchi+13}). The evidence of
systemic rotation of NGC 1904 along the line-of-sight is apparent.
Rotation patterns as clear as that found here are now detected in a
growing number of cases (NGC 4372 in \citealt{kacharov+14}, 47 Tucanae
in \citealt{bellini+17}, M5 by \citealt{lanzoni+18a}, NGC 5986 by
\citealt{lanzoni+18b}, NGC 6362 by \citealt{Dalessandro+21}, see also \citealt{bianchini+13,vasiliev+21}).

\begin{deluxetable*}{rrrrccrrcc}
\tablecaption{Line-of-sight rotation curve of NGC 1904. \label{tab_rotcurve}}
\tablewidth{0pt}
\tablehead{
\colhead{XR$_i$ } & \colhead{ XR$_e$ } & \colhead{ XR$_m+$ } & \colhead{ $N+$}  &  \colhead{$V_{\rm rot}+$}  &
\colhead{ $\epsilon_{V+}$} & \colhead{ XR$_m-$}  & \colhead{ $N-$} & \colhead{$V_{\rm rot}-$} &\colhead{ $\epsilon_{V-}$} 
}
\startdata
    5  &  20  &  10.88 &  139 &   0.20  & 0.49 & $-8.41 $ & 212 & $-0.40$ & 0.49\\
    20 &  50  &  31.99 &  102 &   1.20  & 0.57 & $-33.17$ & 78  & $-1.40$ & 0.61 \\
    50 &  110 &  70.46 &  36  &   1.80  & 0.57 & $-72.49$ & 28  & $-1.00$ & 0.63 \\
    110&  200 & 141.51 &  9   &   0.60  & 0.84 & $-142.11$ & 11  & $-1.40$ & 0.94\\
\enddata
\tablecomments{Rotation curve of NGC 1904.
For four intervals of projected distances from the rotation axis (XR), the table lists: the inner and outer absolute limits of each bin (XR$_i$ and XR$_e$ ) in arcseconds, the mean distance, number of stars, average velocity and its error (in km s$^{-1}$) on the positive side of the XR axis (columns 3–6), and on its negative side (columns 7–10).}
\end{deluxetable*}

\subsection{Velocity dispersion profile}
\label{sec_vd}
\begin{figure}[ht!]
\includegraphics[width=11.6cm, height=9.7cm]{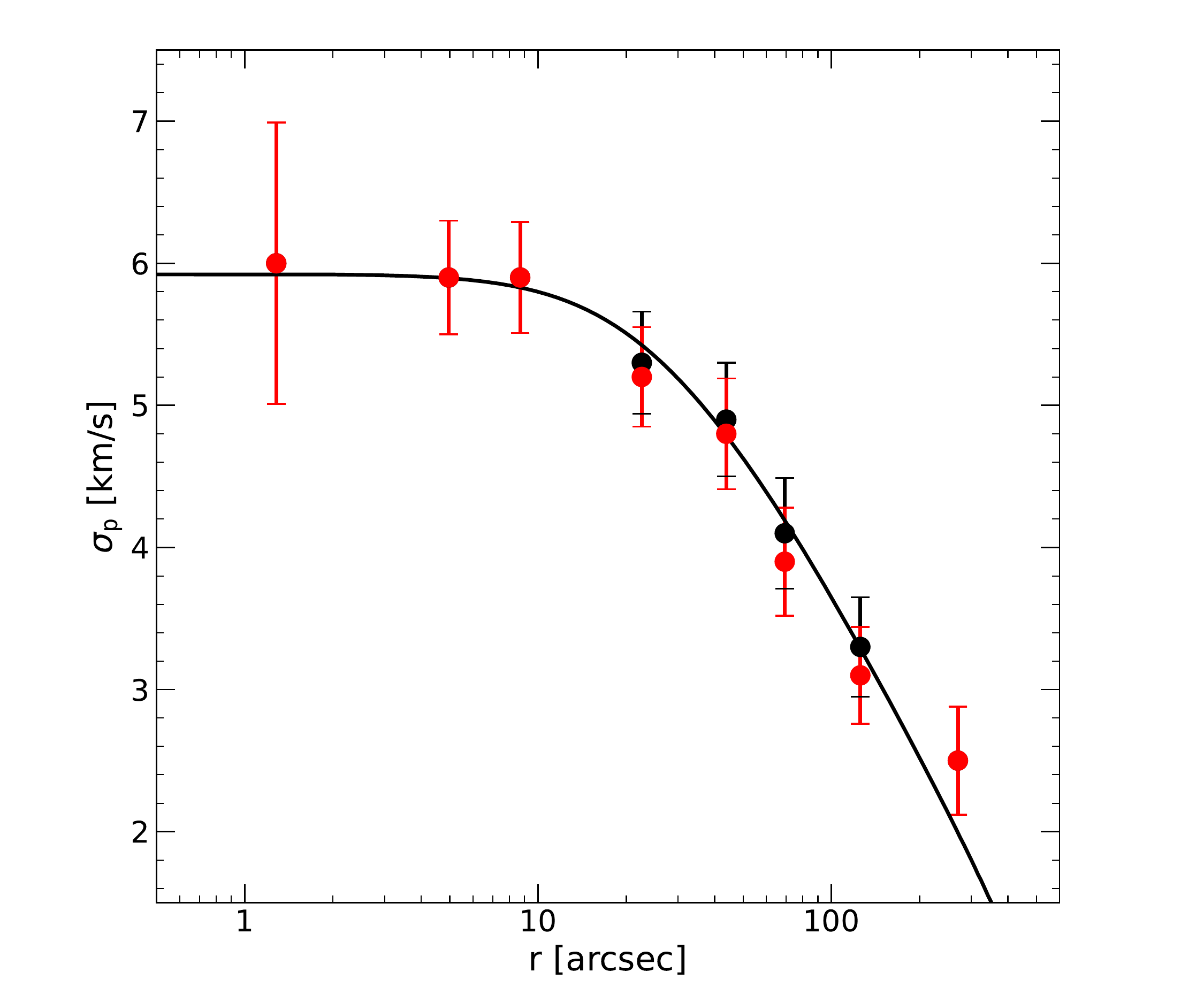}
\centering
\caption{Velocity dispersion profile of NGC 1904 (red circles)
  obtained after subtracting the contribution of rotation in all radial
  bins.  The second velocity moment profile (the same as in Figure
  \ref{fig:2ndv_combi}), which includes the effects of both rotation
  and velocity dispersion in each circular shell, is also shown for
  comparison (black circles). The sold line shows the King model that
best fits the projected density distribution of NGC 1904 by \citet[][]{miocchi+13}.}
\label{fig:fig_vd}
\end{figure}

Once the line-of-sight rotation curve is determined, the projected
velocity dispersion profile, $\sigma_P(r)$, can be finally obtained
from the dispersion of the measured RVs after subtraction of the
ordered motion contribution. We thus assigned to each star the mean
rotational velocity of the XR shell to which it belongs, and
subtracted this value from the measured RV. Then, we repeated the
Maximum-Likelihood procedure described in Section \ref{sec:2ndv} to
determine the projected velocity dispersion profile of NGC 1904 in
circular concentric shells.  The result is shown in Figure \ref{fig:fig_vd}
(red circles) and listed in Table \ref{tab_vdisp}. For the sake of
comparison we also show the radial profile of second velocity moment (black circles)
determined in Section \ref{sec:2ndv} and listed in Table \ref{tab:2ndv_01}.

By construction the velocity dispersion is systematically smaller than
the second velocity moment in every bin. However, the differences are
small and always within the errors, as expected in the case of a
pressure-supported system.  Indeed, in spite of a clean rotation
signal, the rotational velocity is small and it is smaller than the
velocity dispersion in all radial bins, confirming that the kinematics
of NGC 1904 is dominated by non-ordered motions.

\begin{deluxetable*}{LRRRCC}
\tablecaption{Velocity dispersion profile of NGC 1904.}
\tablewidth{0pt}
\tablehead{
\colhead{ $r_i$ }  & \colhead{ $r_e$ }  & \colhead{$r_m$}  &
\colhead{$N$} & \colhead{$\sigma_{P}$} & \colhead{$\epsilon_{\sigma_{P}}$} \\  
\colhead{ [arcsec] } & \colhead{ [arcsec] }  & \colhead{[arcsec]}  &  \colhead{ } &\colhead{km s$^{-1}$ }  & \colhead{ km s$^{-1}$} 
}
\startdata
0.01  &  2.00   &  1.28  &  31 &  6.00 & 0.99  \\
2.00  &   7.00  &  4.96  & 269 &  5.90 & 0.40  \\
7.00  &  11.00  &  8.69  & 259 &  5.90 & 0.39  \\
11.00 &  35.00  &  22.58 & 203 &  5.20 & 0.35  \\
35.00 &  55.00  & 43.84  & 138 &  4.80 & 0.39  \\
55.00 & 100.00  & 69.37  & 73  &  3.90 & 0.38  \\
100.00 & 170.00 & 125.57 & 43  &  3.10 & 0.34  \\
170.00 & 600.00 & 270.13 & 23  &  2.50 & 0.38  \\
\enddata
\tablecomments{The first four columns list the internal, external, mean radii and number of stars of each adopted radial bin ($r_i$, $r_e$, $r_m$ and $N$, respectively). The last two columns list the velocity dispersion and its uncertainty in each bin, respectively.}
\label{tab_vdisp}
\end{deluxetable*}

\subsection{Rotation in the Plane of the Sky}
\label{sec_rotPM}
\begin{figure}[!h]
\includegraphics[width=17cm, height=7.7cm]{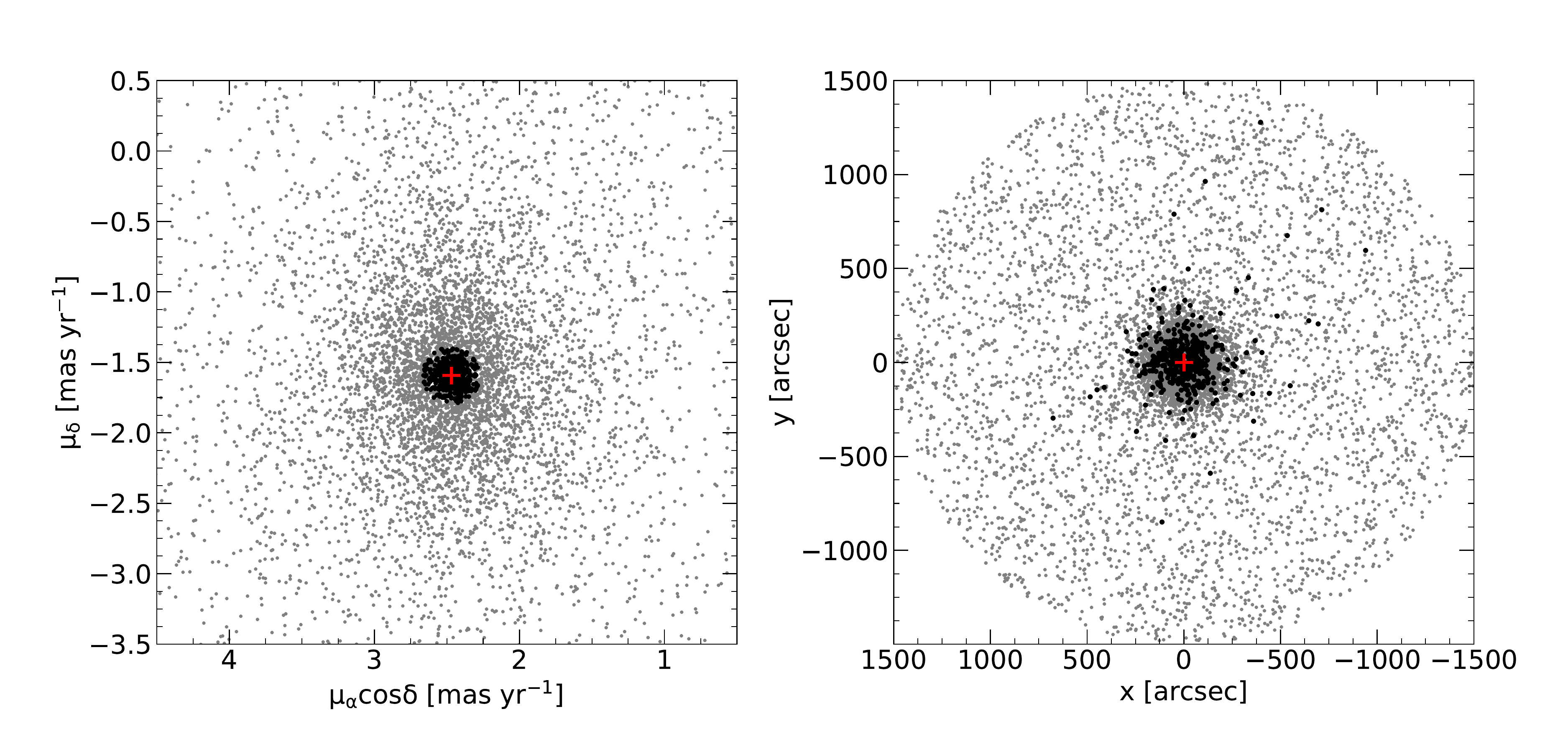}
\centering
\caption{{\it Left panel}: VPD of the stars of the selected sample (black circles) and of the original Gaia EDR3 dataset (gray points), with the red cross marking the absolute motion \citep{vasiliev+21}. {\it Right panel}: Map on the plane of the sky, with respect to the adopted cluster center (red cross), of the selected star sample (black circles) and of the original Gaia EDR3 one (gray points).\label{fig:fig_pm}}
\end{figure}

In order to study the rotation of the cluster in the plane of the sky
we used the most recent data release (EDR3) of the Gaia mission
\citep{gaia+21}. Using this dataset \citet{vasiliev+21} have detected
a signal of rotation in NGC 1904 at more than $3\sigma$ confidence level and with a maximum amplitude of $\sim2.0$ km s$^{-1}$ at $\sim2\arcmin$. In a previous analysis using both line-of-sight velocities and proper motions (PMs) from Gaia DR2, \citet[][]{sollima+19} quoted a rotation velocity amplitude of $2.24 \pm 0.46$ km s$^{-1}$, but classified the rotation signal as uncertain, because the cluster failed one of the tests performed against random or systematic effects (see their Table 1).
Here we independently analyzed the Gaia EDR3 data searching for signature of rotation in the plane of the sky. We first applied the following selection criteria on the astrometric and photometric quality indicators (given in the Gaia archive and discussed in \citealp{Lindegren+21,Riello+21,Fabricius_2021}), in order to use the stars with the most reliable astrometric measures: ($i$) RUWE\footnote{RUWE is the renormalised unit weight error (for astrometry) discussed in \citet{Lindegren+21}.} $< 2.4$; ($ii$) astrometric\_excess\_noise\footnote{it is the excess noise of the source and it measures the disagreement, expressed as an angle, between the observations of a source and the best-fitting standard astrometric model.} $< 2$; ($iii$) phot\_bp\_rp\_excess\_factor\footnote{it is the sum of the integrated BP and RP fluxes divided by the G flux. It gives an indication of the consistency between the three fluxes \citep[for more details see][]{Riello+21}.} $< 2.6 + 0.12 (\mathrm{BP}-\mathrm{RP})^{2}$, where BP and RP are the magnitude in the BP and RP bands, respectively;
($iv$) G $<19$; 
($v$) PM errors $<0.06$ mas yr$^{-1}$ for stars with G $<16$, while for G $>16$ we have divided the sample into bins of distance (0\arcsec - 50\arcsec, 50\arcsec - 100\arcsec, 100\arcsec - 150\arcsec and distance $>150\arcsec$) and bins of magnitude 0.5 G magnitude wide, and we have excluded the stars with PM errors larger than $1\sigma$ of the local mean error in each bin. Finally, we selected stars within 0.2 mas yr$^{-1}$ (corresponding to about $2\times \sigma_0$ at the distance of NGC 1904) from the absolute motion \citep{vasiliev+21} in the vector-point diagram (VPD, see the left panel of Figure \ref{fig:fig_pm}).
The final sample includes 437 stars located between 7\arcsec \ and 1338\arcsec \ from the center. Their $\mu_{\alpha}cos\delta$ and $\mu_{\delta}$ PMs are shown (black dots) in the VPD in the left panel of Figure \ref{fig:fig_pm}, while their distribution on the plane of the sky is plotted in the right panel of the same figure. The PMs measured for the selected stars and their corresponding uncertainties have been converted into a Cartesian reference frame centered on the cluster center using eq. 2 of \citet{gaia2018} and corrected for perspective effect using eq. 6 of \citet{G_van_de_Ven_2005} and assuming a cluster distance of 13.2 kpc \citep[][consistent with the more recent estimate by \citealp{Baumgardt+21}]{ferraro+99}. PMs are then decomposed into projected tangential $\mu_{TAN}$ and radial $\mu_{RAD}$ components and converted into units of km s$^{-1}$ using eq. 4 of \citet{G_van_de_Ven_2005}.

\begin{figure}[ht!]
\includegraphics[width=15cm, height=16cm]{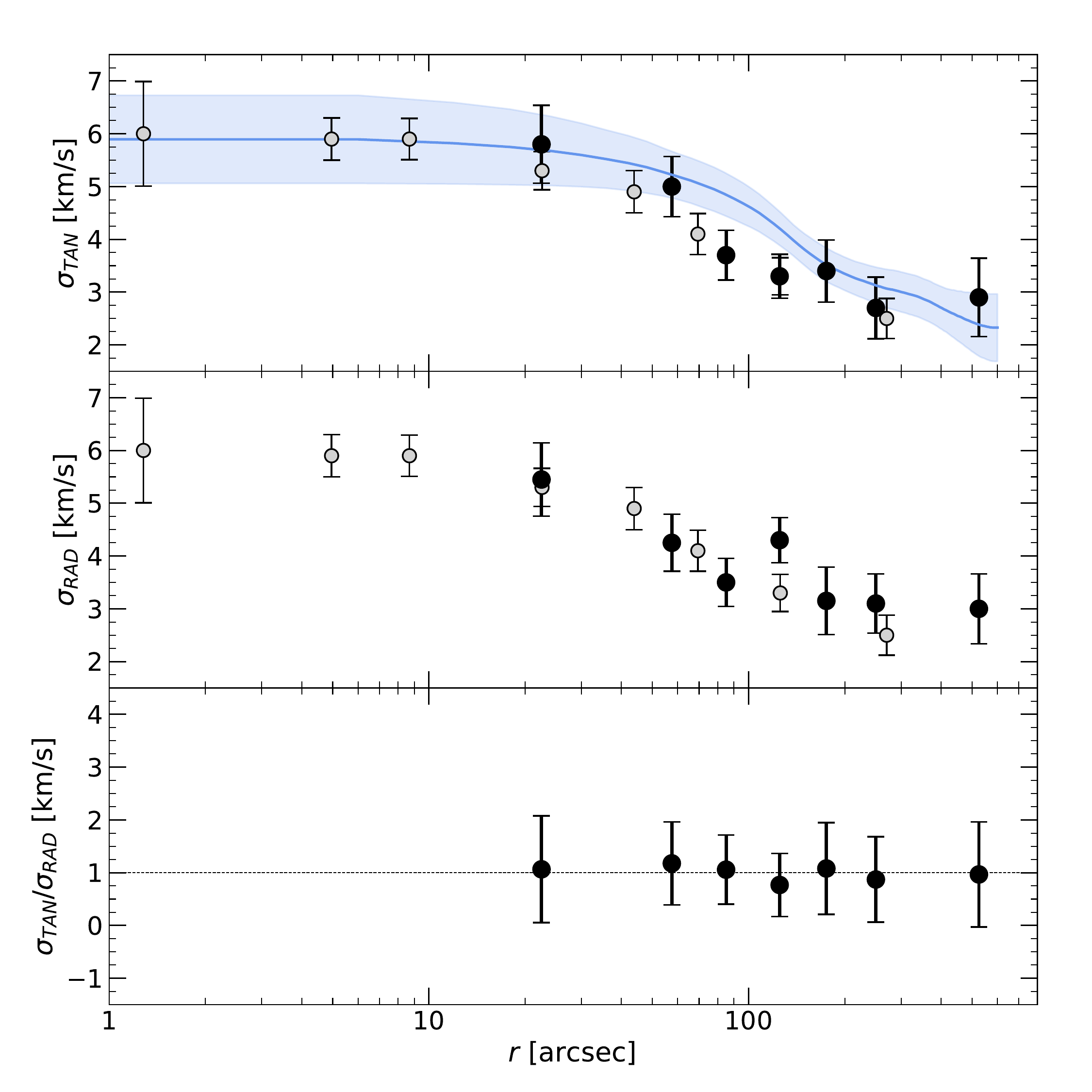}
\centering
\caption{{\it From top to bottom}: Tangential and radial velocity dispersion profiles, and anisotropy profile. The black circles mark the results obtained with the sample selected from the Gaia EDR3 dataset, while the grey circles correspond to the second velocity moment profile obtained in this work from individual RVs (the same as in Figure \ref{fig:2ndv_combi}). The PM dispersion profile presented in \citet{vasiliev+21} is marked for comparison by the blue solid line in the top panel, with shaded area representing its 68\% confidence interval.} \label{fig:fig_kin_pm_01}
\end{figure}

\begin{figure}[ht!]
\includegraphics[width=13.8cm, height=10.8cm]{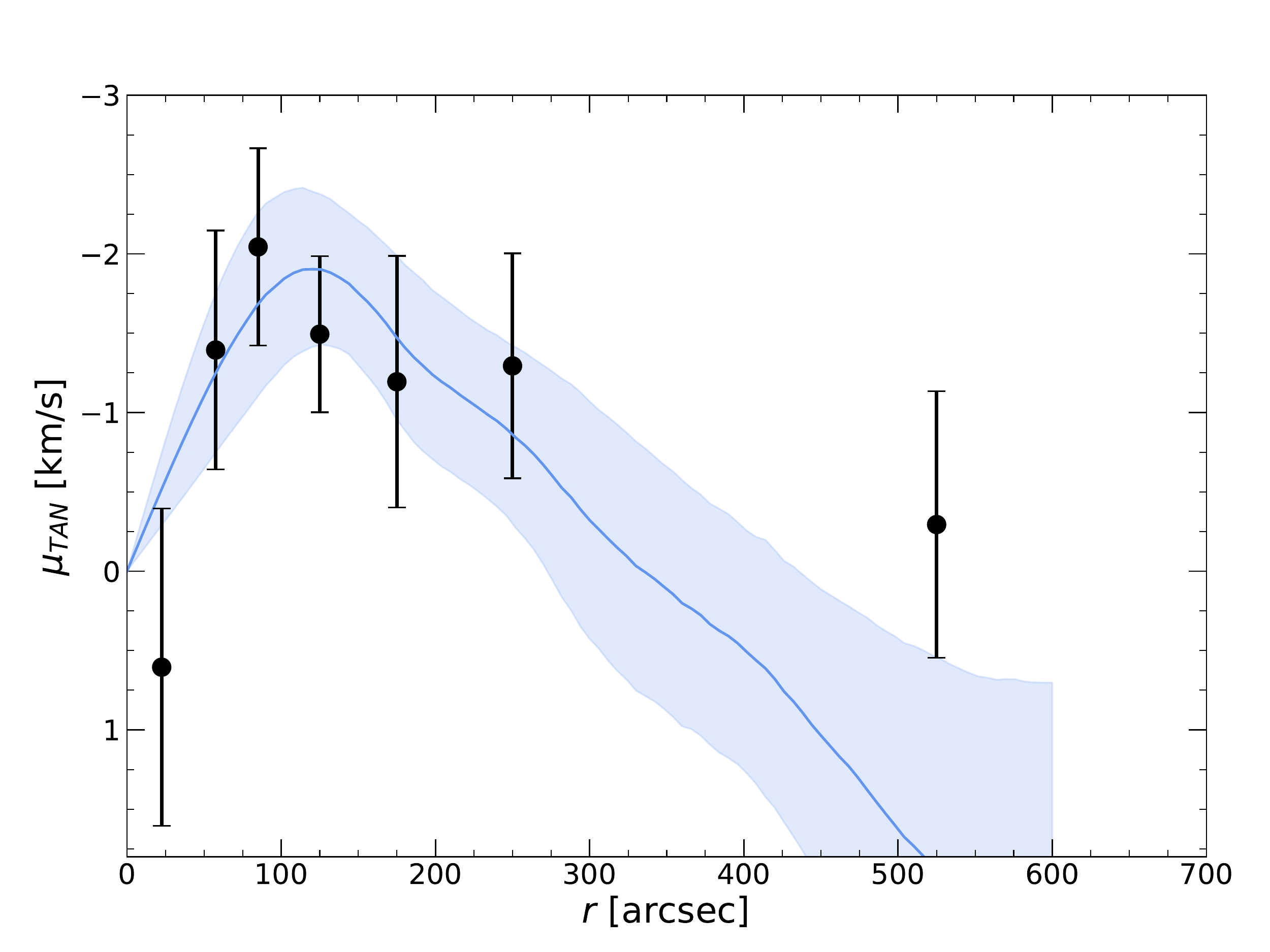}
\centering
\caption{Rotation profile of NGC 1904 in the plane of the sky obtained with the star sample selected in this work from the Gaia EDR3 data (black circles). The solid blue line and shaded area show the PM rotation profile presented in \citet[][]{vasiliev+21} and its 68\% confidence interval, respectively.} \label{fig:fig_kin_pm_02}
\end{figure}

Similarly to what has been done for the line-of-sight component, we derived the velocity dispersion in both the tangential ($\sigma_{TAN}$) and radial ($\sigma_{RAD}$) components by dividing the sample in radial bins and using a Maximum-Likelihood approach. The results are listed in Table \ref{tab_kin_pm} and shown in Figure \ref{fig:fig_kin_pm_01} (top and central panels).
For comparison purposes, we also show the PM dispersion profile obtained by \citet[][]{vasiliev+21} using the Gaia EDR3 dataset (blue solid line in the top panel of Figure \ref{fig:fig_kin_pm_01}). Within the errors, their results are in agreement with the profile obtained in this work (black circles), except for a couple of points at $r \sim 100\arcsec$.
It is interesting to note that the radial profile of $\sigma_{TAN}/\sigma_{RAD}$ shown in the bottom panel of Figure \ref{fig:fig_kin_pm_01} indicates that this cluster is characterized by an isotropic velocity distribution, although the large uncertainties do not allow to rule out the possible presence of anisotropy. As shown in \citet{tiongco+16} an isotropic  velocity distribution would suggest that this cluster is in an advanced dynamical stage and lost any velocity anisotropy developed during its early or long-term phases of evolution (see Section \ref{sec_discuss} below for further discussion).
We also derived the rotation curve in the plane of the sky by computing, again with a Maximum-Likelihood method, the average value of $\mu_{TAN}$ in each radial bin. We assumed negative $\mu_{TAN}$ values in the case of a clockwise rotation in the plane of the sky (from North to West). The obtained rotation profile is shown in Figure \ref{fig:fig_kin_pm_02} (black circles) and listed in Table \ref{tab_kin_pm} (last two columns). In Figure \ref{fig:fig_kin_pm_02}, the rotation curve is compared with that published in \citet[][blue solid line]{vasiliev+21}, showing a quite good agreement. In agreement with what derived from line-of-sight measurements, we find a signature of clockwise rotation even in the plane of the sky, with a maximum amplitude of $\sim-2.0$ km s$^{-1}$ at $\sim80$\arcsec \ ($2$ projected half-mass radii or $\sim 1.5$ three-dimensional half-mass radius) from the cluster center.
As a final result, the rotation profile both in the plane of the sky and in the line-of-sight direction will be compared with rotation curve models with different inclination angles in Section \ref{sec_discuss}.
 
\begin{deluxetable*}{rrrccccrc}[!ht]
\tablecaption{Radial and tangential velocity dispersion profile and rotation curve in the plane of the sky of NGC 1904. \label{tab_kin_pm}}
\tablewidth{0pt}
\tablehead{
\colhead{$r_i$ } & \colhead{ $r_e$ } & \colhead{ $N$ } & \colhead{ $\sigma_{RAD}$}  & \colhead{ $\epsilon_{\sigma_{RAD}}$} & \colhead{ $\sigma_{TAN}$} & \colhead{ $\epsilon_{\sigma_{TAN}}$} & \colhead{$\mu_{TAN}$} &\colhead{ $\epsilon_{\mu_{TAN}}$}  \\
\colhead{ [arcsec] } & \colhead{ [arcsec] }  &  \colhead{ } &\colhead{km s$^{-1}$ } &\colhead{km s$^{-1}$ } &\colhead{km s$^{-1}$ } &\colhead{km s$^{-1}$ } &\colhead{km s$^{-1}$ } &\colhead{km s$^{-1}$ }
}
\startdata
    5  &  45  &  38 &  5.45 & 0.69 &  5.80 & 0.74 &  0.61    & 1.00  \\
    45 &  70  &  65 &  4.25 & 0.54  & 5.00 & 0.57 & $-1.39$ & 0.75\\
    70 &  100 &  79 &  3.50 & 0.45 & 3.70 & 0.47  & $-2.04$ & 0.62 \\
    100 & 150 &  104 & 4.30 & 0.43 & 3.30 & 0.42 & $-1.49$ & 0.49 \\
    150 & 200 &  52  & 3.15 & 0.64 & 3.40 & 0.59 & $-1.19$ & 0.79 \\
    200 & 300 &  59 &  3.10 & 0.56 & 2.70 & 0.59 & $-1.29$ & 0.71 \\
    300 & 750 &  33 &  3.00 & 0.66 & 2.90 & 0.74 & $-0.29$ & 0.84\\
\enddata
\tablecomments{The table lists: the internal and external radii of each adopted radial bin ($r_i$ and $r_e$, respectively), the number of stars in each bin ($N$), the velocity dispersion and its uncertainty in the radial and tangential component (columns 4–7) and the average $\mu_{TAN}$ and its uncertainty in each radial bin (columns 8–9).}
\end{deluxetable*}

\begin{figure}[ht!]
\includegraphics[width=12cm, height=12cm]{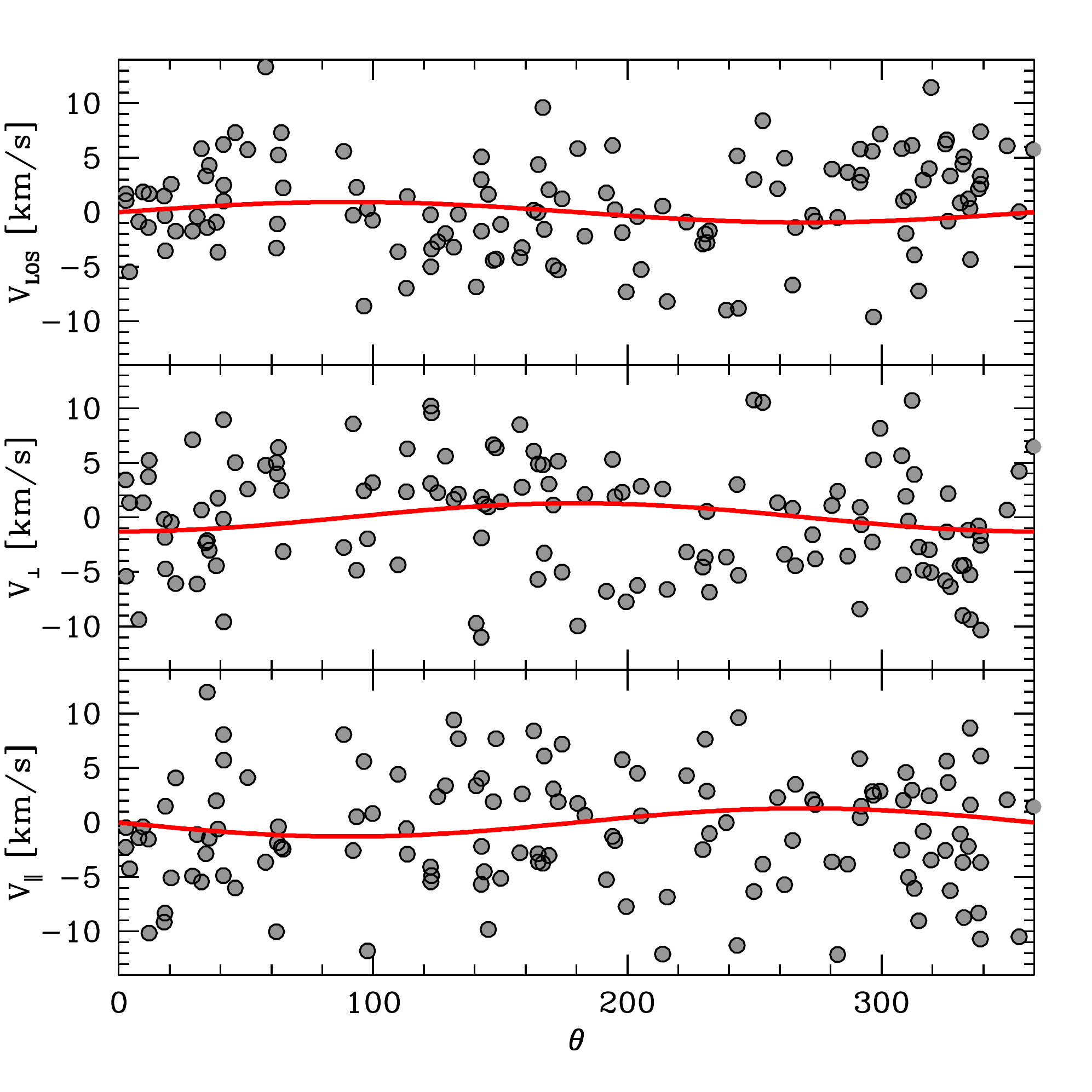}
\centering
\caption{Distribution of the three velocity components for the 130
  stars in NGC 1904, for which the three-dimensional kinematic information is available, as a function of their angular position in degree. The red solid lines
  show the curves obtained for PA$_0 = 98 \arcdeg$ and for the best-fit values of the rotation amplitude $A=-2.05$ km s$^{-1}$ and inclination angle $i=37\arcdeg$, as discussed in the text (see Section \ref{sec_rotPM}).} \label{fig:fig_3d}
\end{figure}

Since for a sub-sample of 130 stars we have the three velocity components (i.e. RV, $\mu_{\alpha}cos\delta$ and $\mu_{\delta}$), as sanity check, we applied the method described in \citet{sollima+19} to verify that the values of PA$_0$ and rotation amplitude estimated separately along the line of sight and the plane of the sky directions are able to properly reproduce also the three-dimensional velocity space of the system as traced by this sub-sample. Following \citet{sollima+19}, in case of rotation, a modulation of
the mean velocity in the three components as a function of the angular position of the stars is detected. According to their eq. (2), for a solid-body rotation the three velocity components are:
\begin{eqnarray}
   \rm RV \ & = & \ \omega \ R \ sin \ (\rm{\theta} - \rm{PA}_0) \ \textit{sin i}  \nonumber \\
   v_{\parallel} \ & = & \ \omega \ R \ sin \ (\rm{\theta} - \rm{PA}_0) \ \textit{cos i} \\
    v_{\bot} \ & = &  \ \omega \ R  \ cos \ \rm (\theta- PA_0) \ \textit{cos i} \nonumber 
\end{eqnarray}
where $\omega$ is the angular velocity, {\it R} is the projected distance from the cluster center, $0\arcdeg< \theta < 360 \arcdeg$ is the angular position of the stars (growing anti-clockwise from North to Est), $0\arcdeg< i < 90 \arcdeg$ is the inclination angle of the rotation axis with respect to the line-of-sight, 
PA$_0$ is the position angle of the rotation axis defined as above, RV is the velocity component along the line-of-sight and $v_{\parallel}$ and $v_{\bot}$ are the components
in the directions parallel and perpendicular to the rotation axis, respectively \citep[for more details see Appendix A of][]{sollima+19}. 
We then defined $A = \langle \omega \ R \rangle $ as the average projected rotation velocity amplitude and we assumed it negative for clockwise rotation in the plane of the sky.
In our analysis, we have fixed PA$_0$ to the value obtained above from the line-of-sight velocities (PA$_0 = 98 \arcdeg$, see Section \ref{sec_vrot}) and we then used a Maximum-Likelihood algorithm to determine the best-fit values of $i$ ($i=37\arcdeg$) and $A$ ($A=-2.05$ km s$^{-1}$). The latter is in very good agreement with the estimates obtained from the previous analyses along the line-of-sight and on the plane of the sky, separately. The result of this analysis is plotted in Figure \ref{fig:fig_3d}.

\newpage
\subsection{Ellipticity}
\label{sec_ellipticity}
\begin{figure}[ht!]
\includegraphics[width=10.8cm, height=9.2cm]{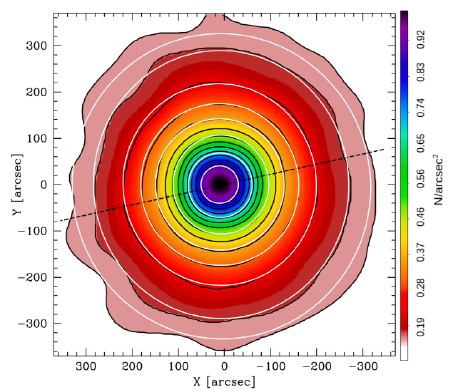}
\centering
\caption{Smoothed stellar density map of the inner $400\arcsec \times 400\arcsec$ of NGC 1904, obtained from the photometric catalog discussed in \citet[][]{lanzoni+07}. The solid black lines are isodensity contours, while the white curves mark their best-fit ellipses. The black dashed line marks the direction of the global rotation axis (with position angle of $98\arcdeg$, see Section \ref{sec_kin}). \label{fig:fig_density}}
\end{figure}

A rotating system is also expected to be flattened in the direction perpendicular to the
rotation axis \citep{Chandrasekhar_1969}. Although the cluster rotation measured along the line-of-sight direction has a maximum amplitude of just $\sim 1.5$ km s$^{-1}$, we investigated the morphology of NGC 1904 by using the catalog discussed in \citet[][]{lanzoni+07} to build the two-dimensional (2D) stellar density map of the system. The analysis was limited to an area of $\sim 400\arcsec \times 400\arcsec$ and only stars with $V<21$ have been used to avoid incompleteness effects. The distribution of star positions was transformed into a smoothed 2D surface density function through the use of a kernel with a width of $1\arcmin$ \citep[see][]{dalessandro+15}.
The resulting 2D density map is shown in Figure \ref{fig:fig_density}, where the black solid lines represent the isodensity curves, the white lines correspond to their best-fit ellipses and the dashed line marks the position of the rotation axis estimated from the star RVs (PA$_0 = 98 \arcdeg$).
As is clear from the figure, the stellar density distribution has spherical symmetry in the center and becomes slightly more flattened for increasing radius.
The ellipticity $\epsilon=1-b/a$ where $a$ and $b$ are the major and minor axis, respectively, reaches its maximum in the external region (at $r\sim220\arcsec$) where we measured a value of $ 0.04\pm0.02$, qualitatively consistent with what found previously by \citet[][$0.01$]{harris96} and in line with the small measured rotation. Although the ellipticity is small, ellipses major axis tend to have an orientation of $\sim 10\arcdeg$ in the North-East
direction, implying that the stellar density distribution is flattened in the direction perpendicular to the rotation axis in agreement with what is expected for a rotating system and qualitatively consistent with that predicted, for example, by the models introduced by \citet{varri+12} and found in other observational studies (e.g., \citealp{Dalessandro+21}; see also \citealp{bianchini+13, bellini+17, lanzoni+18a}).

\section{Discussion}
\label{sec_discuss}
In the context of the ongoing MIKiS survey
\citep{ferraro+18b}, in this paper, we presented the velocity
dispersion profile and rotation curve of the GGC NGC 1904. Thanks to a
combination of different datasets acquired with appropriate spatial
resolution, we measured the RV of $\sim 1700$ individual stars
sampling the entire cluster radial extension.

Figure \ref{fig:fig_vd} shows that the velocity dispersion profile of NGC 1904 is characterized by approximately constant behaviour in the innermost region and a monotonically decreasing one in the outer part. We can compare this behavior to
the expectation from the \citet{king66} model that best fits the
observed star density distribution, to verify whether it is able to
simultaneously reproduce the observed structure and kinematics.
In doing this, we always prefer to use the star density profile from
resolved star counts, instead of the surface brightness profile, to
avoid possible biases due to the sparse presence of bright stars,
which can locally alter the luminosity distribution. As discussed in
\citet{miocchi+13}, the King model that best fits the star density
profile of NGC 1904 is characterized by a dimensionless central
potential $W_0=7.75$, which corresponds to a concentration parameter
$c=1.76$, defined as $c=\log(r_t/r_{\rm 0})$, where $r_t=9.32\arcmin$ is the
cluster tidal radius and $r_{\rm 0}=9.8\arcsec$ is the King radius, which is similar,
although not coincident, with the core radius $r_{\rm c}=9.4\arcsec$. The
three-dimensional half-mass radius and the projected half-mass radius of the system 
are $r_{\rm h}=56.7\arcsec$ and $R_{\rm h}=41.7\arcsec$, respectively.
Adopting the distance modulus quoted in \citet{ferraro+99}, which
locates the cluster at a distance of $\sim 13.2$ kpc from us, the
physical size of the cluster structural parameters turn out to be: $r_{\rm c}= 0.6$ pc
and $r_{\rm h}=3.63$ pc.  We adopted these values as representative of the
structure of the cluster, and we thus compared the projected velocity dispersion 
profile of this model to the observations 
(see Figure \ref{fig:fig_vd}). By performing a $\chi^2$ test, we found the value of $\sigma_0$ that minimizes the residuals between the observed velocity dispersion profile and the adopted King model. From the solutions providing $\chi^2 = \chi^{2}_{min} \pm 1$, we then obtained the $1\sigma$ uncertainty on the best fit value. The resulting central velocity dispersion is $\sigma_0 =5.9 \pm 0.3$ km s$^{-1}$, and the comparison with the observed profile clearly shows that the
adopted King model is able to nicely reproduce both the observed
structure and the observed kinematics of NGC 1904.
This value is in good agreement with previous
determinations: it is only slightly larger than the value listed in
the \citet[][]{harris96} catalog ($\sigma_0 = 5.3$ km s$^{-1}$), and slightly
smaller than that found by \citet{Baumgardt+18}, 
who quote $\sigma_0 =6.5$ km s$^{-1}$, in spite of the fact that they 
adopt  the velocity dispersion profile by \citet{lutz+13} who, conversely obtain a significantly larger value of the central velocity dispersion ($\sigma_0 >8$ km s$^{-1}$).

From the derived value of the central velocity dispersion we estimated the total mass of the system. To this end, we used equation (3) in
\citet{majewski+03}, where the parameters $\mu$ and $\beta$ have been
determined, respectively, by following \citet{djorgovski93} and by
assuming $\beta=1/\sigma^2_0$ (as appropriate for models with $W_0>5$; see the discussion in \citealt{richstone+86}). 
The resulting total mass is $M=1.28_{-0.14}^{+0.15} \times 10^5 M_\odot$. The uncertainties have been estimated through a Monte Carlo simulation, performing 1000 random extractions of the values of $c$, $r_0$ and $\sigma_0$ from normal distributions centered in their best fit values and with dispersions equal to the estimated uncertainties of the respective parameters.
This mass estimate is in perfect agreement with the value obtained in \citet[][$1.29 \times 10^5 M_\odot$]{ferraro+18b}, and slightly smaller than the values quoted in \citet[][$1.58_{-0.14}^{+0.15}\times 10^5 M_\odot$]{mc+05}, \citet[][$1.44 \pm 0.10 \times 10^5 M_\odot$]{lutz+13}, and \citet[][$1.69 \pm 0.11 \times 10^5 M_\odot$]{Baumgardt+18}. These differences can be justified taking into account all the uncertainties and the fact that the estimates have been obtained through different methods.

A signature of systemic rotation in the outskirts of this cluster
($85\arcsec <r <200\arcsec$), with a maximum amplitude of 1.7 km
s$^{-1}$ and a position angle of $108\arcdeg$ for the rotation axis,
was already presented in \citet{ferraro+18b}.
Here we unambiguously confirm the presence of global rotation in the line-of-sight component using complementary data that sample the innermost regions.  
A well-defined rotation pattern with a rotation axis position angle of $98\arcdeg$ has been found.
The resulting rotation curve is presented in Figure \ref{fig:fig_rotcurve} and is well described by the analytical expression introduced in \citet{lyndenbell67} to describe the equilibrium rotational profile emerging at the end of violent relaxation. The peak of the rotation curve ($A_{\rm peak} \sim 1.5$ km s$^{-1}$) is located at a distance from the cluster center of about $1.5$ three-dimensional half mass radius (or nearly 2 projected half-mass radii). 
To obtain the full characterization of the internal kinematics of the cluster, we analyzed the stellar PMs from the Gaia EDR3 data, detecting a signal of rotation also in the plane of the sky (see Figure \ref{fig:fig_kin_pm_02}). To double check these results, we then applied the procedure described in \citet{sollima+19} to the sub-sample of 130 stars for which we have the three velocity components (RVs and PMs). By fixing the rotation axis position angle  to the value obtained from the measured RVs ( PA$_0=98 \arcdeg$), this analysis confirmed the presence of a rotation pattern with the previously determined amplitude, and it provided us with an estimate of the inclination angle of the rotation axis with respect to the line-of-sight: $i=37$\arcdeg.
The main final results are summarized in Table \ref{tab_final}, which also lists all the parameters of the cluster used in this work.

\begin{figure}[ht!]
\includegraphics[width=16cm, height=12cm]{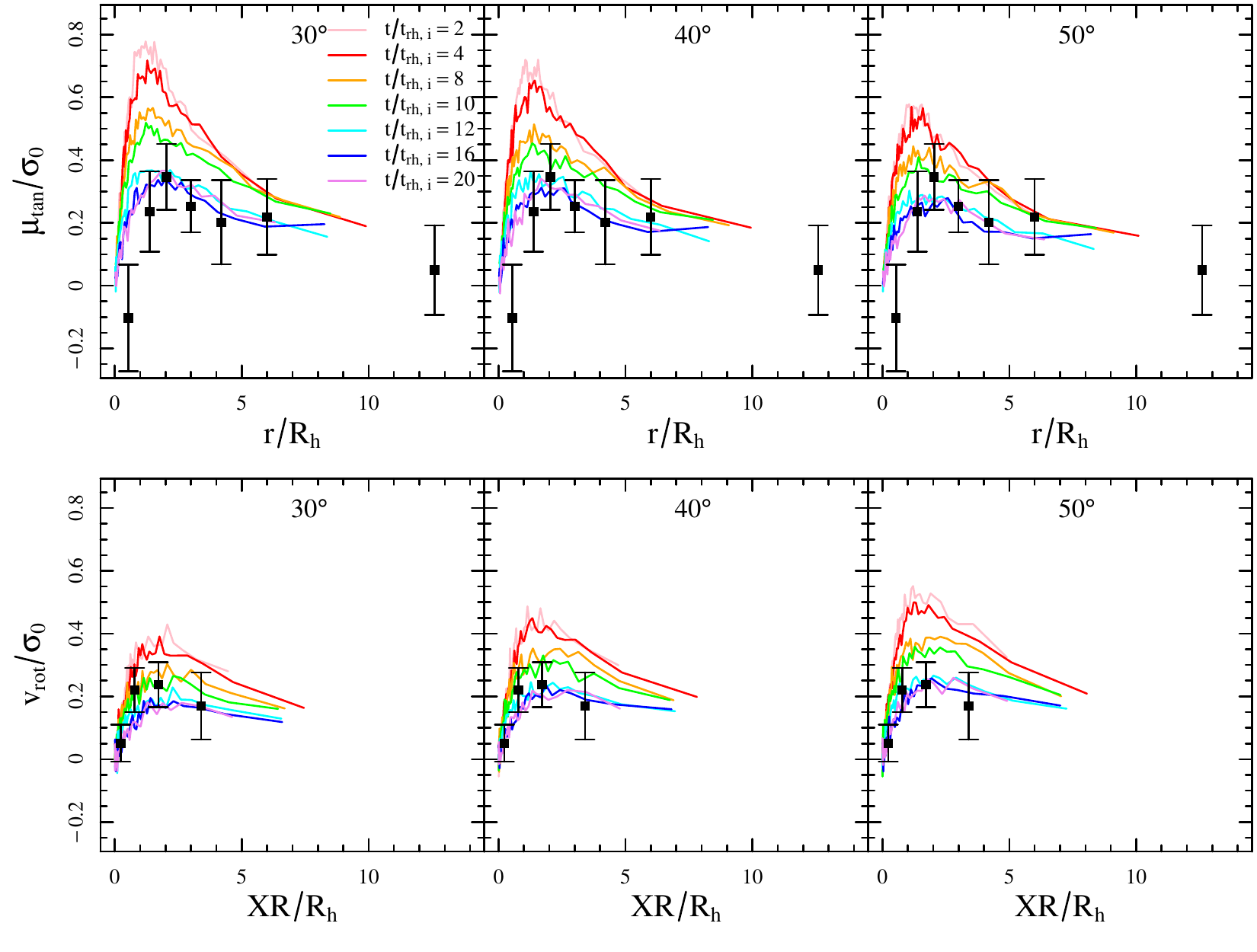}
\centering
\caption{Comparison between the observed rotation curves normalized to the central velocity dispersion (black squares) and the results from the {\it N}-body simulations of \citet[][VBrotF04 model, see their Table 1]{tiongco+16}, obtained at different evolutionary times (curves of different colors; see labels in the top-left panel, where $t_{hr,i}$ is the cluster's initial half-mass relaxation time). The top and bottom panels show, respectively, the rotation in the plane of the sky and that in the line-of-sight direction. From left to right, the results obtained for different inclination angles between the line-of-sight and the rotation axis of the simulated cluster are shown: $i=30 \arcdeg$, $i=40 \arcdeg$, $i=50 \arcdeg$, as labelled. The distance from the cluster center (top panel) and the distance from the rotation axis (bottom panel) are normalized
to the projected half-mass radius ($R_{\rm h}$).
  \label{fig:fig_simu}}
\end{figure}

Theoretical studies \citep[e.g.,][]{fiestas+06, Ernst2007, hong2013, tiongco+17} have shown that star clusters gradually lose their initial rotation as a result of the effects of internal two-body relaxation and angular momentum loss carried away by escaping stars. Hence, the extent of the present-day rotation detected in many GGCs \citep[see, e.g.,][]{ bianchini+13,fabricius+14,bellini+17, ferraro+18b, kamann+18, lanzoni+18a, sollima+19, Dalessandro+21} represents only a lower limit to the primordial rotation of these systems and suggests that the initial cluster kinematics might be characterized by a stronger rotation
\citep{HenaultBrunet+12,Mackey_2012,kamann+18b, Dalessandro+21b}.
To address this issue for the case of NGC 1904, in Figure \ref{fig:fig_simu} we compare the observed rotation (black squares) with the results (colored lines) of a representative $N$-body simulation of a rotating star cluster from the survey of models presented in \citet{tiongco+16}. The results reported here are those from the VBrotF04 model (see \citealt{tiongco+16} for further details), a system initially set up with phase space properties following those of the rotating models introduced by \citet{varri+12}. We emphasize that a study specifically aimed at identifying the best evolutionary model for NGC 1904 would require an extensive investigation following the evolution of systems with a broad range of different initial conditions. The purpose of our comparison here is just to illustrate a general dynamical path leading to the observed rotational properties of NGC 1904. Figure \ref{fig:fig_simu} shows the rotation curves in the plane-of-the-sky (upper panels), and in the line-of-sight direction (lower panels). From left to right, the inclination angle of the simulated cluster rotation axis with respect to the line-of-sight direction varies from $i=30 \arcdeg$, to $i=40 \arcdeg$, to $i=50 \arcdeg$ (see labels), which are values close to that determined by the analysis presented in Section \ref{sec_rotPM} ($i=37 \arcdeg$). Curves of different colors correspond to simulated rotation curves obtained at different evolutionary times, as labelled in the top-left panel. The comparison with the observations shows a general good agreement, both on the plane of the sky and along the line-of-sight direction, with the rotation curves obtained in advanced evolutionary phases of the $N$-body model,  when the system has lost a significant fraction of its initial angular momentum. Hence, the cluster rotational properties suggest an old dynamical age for NGC 1904, in agreement with what inferred from the radial distribution of blue straggler stars (see \citealp{ferraro+12,ferraro+18b, ferraro+20,lanzoni+16}). 
This study shows the importance of a complete three-dimensional kinematic characterization of stellar systems. Indeed, it allows not only a proper determination  of the actual strength of the cluster internal rotation, but also to constrain the dynamical phase of the system. The synergy between measurements of the structural and kinematic properties, as in the case of NGC 1904, can provide the essential ingredients necessary to build a complete picture of the formation and dynamical history of GGCs.

\begin{deluxetable*}{lCc}
\tablecaption{Summary of the parameters used and the main results obtained in this work for the GGC NGC 1904. \label{tab_final}}
\tablewidth{0pt}
\renewcommand{\arraystretch}{1.2}
\tablehead{
\colhead{Parameter} & \colhead{Estimated Value} & \colhead{ Reference}
}
\startdata
    Cluster center [deg] & \rm RA=81.0462112  & \citet{lanzoni+07} \\
                         & \rm Dec=-24.5247211 &   \\
    Color excess & \rm{E}(\textit{B}$-$\textit{V}) = 0.01 &  \citet{ferraro+99}\\
    Metallicity & $\rm [Fe/H]$=-1.6&  \citet{harris96} \\
    Cluster distance & d=13.2 \ \rm kpc &  \citet{ferraro+99} \\
    3D half-mass radius & $r_{\rm h}$ = $56.7_{-0.005}^{+0}$ &  \citet{miocchi+13} \\
    2D Projected half-mass radius & $R_{\rm h}$ = $41.68_{-0.03}^{+0.08}$ &  \citet{miocchi+13} \\
    Dimensionless central potential  & $\rm W_0$ = $7.75_{-0.1}^{+0.05}$ &  \citet{miocchi+13} \\
    Concentration parameter & $c$ = $1.76_{-0.03}^{+0.02} $ &  \citet{miocchi+13} \\
    Core radius & $r_{\rm c}$ = $9.4_{-0.3}^{+0.6}$ & \citet{miocchi+13}  \\
    Tidal radius & $r_t$ = $9.32_{-0.09}^{+0.04}$ &  \citet{miocchi+13} \\
    Systemic velocity & $V_{sys}$ = 205.4 \pm 0.2 \ \rm km \ s$^{-1}$ & this work \\
    Central velocity dispersion & $\sigma_0$ = 5.9 \pm 0.3 \ \rm km \ s$^{-1}$ & this work  \\
    Line-of-sight rotation peak & $A_{\rm peak}$ = 1.5 \ \rm km \ s$^{-1}$ & this work \\
    Plane-of-the-sky rotation peak &  $A = -2.0$ \ \rm km \ s$^{-1}$ & this work \\
    Rotation axis position angle & PA = 98\arcdeg &  this work\\
    Inclination angle of the rotation axis & $i$ = 37\arcdeg & this work \\
    Absolute Proper Motions & $\mu_\alpha cos\delta$ = 2.469 \pm 0.025 \ \rm mas \ yr$^{-1}$ &   \citet{vasiliev+21} \\
                         & $\mu_\delta$= -1.594 \pm 0.025 \ \rm mas \ yr$^{-1}$ \\
    Ellipticity & $\epsilon$= 0.04 \pm 0.02 &  this work \\
    Total Mass & $M=1.28_{-0.14}^{+0.15}\times10^5 M_\odot$ &  this work \\
\enddata
\tablecomments{All radii are in units of arcseconds, with the exception of $r_t$, which is in arcminutes.}
\end{deluxetable*}

\pagebreak

\newpage
\vskip1truecm
This work is part of the project Cosmic-Lab at the Physics and
Astronomy Department ``A. Righi" of the Bologna University
(http://www.cosmic-lab.eu/ Cosmic-Lab/Home.html). The research was
funded by the MIUR throughout the PRIN-2017 grant awarded to the
project Light-on-Dark (PI:Ferraro) through contract PRIN-2017K7REXT.

ALV acknowledges support from a UKRI Future Leaders Fellowship (MR/S018859/1).


\pagebreak

\bibliography{1904}{}
\bibliographystyle{aasjournal}



\end{document}